\documentclass[11pt]{article}

\RequirePackage{amsthm,amsmath,amsfonts,amssymb}
\RequirePackage[numbers]{natbib}
\RequirePackage[colorlinks,citecolor=blue,urlcolor=blue,backref=page,backref=page]{hyperref}
\RequirePackage{graphicx}
\RequirePackage{float}
\usepackage{tikz}
\usepackage[ruled,vlined]{algorithm2e}
\usetikzlibrary{3d, calc,decorations.markings}
\usepackage[margin=1in]{geometry}
\usepackage{setspace}
\usepackage{algorithmic}

\newtheorem{proposition}{Proposition}[section]

\theoremstyle{plain}

\newtheorem{theorem}{Theorem}[section]

\theoremstyle{definition}
\newtheorem{definition}[theorem]{Definition}

\title{Geodesic Variational Bayes for Multiway Covariances}
\author{
    Quinn Simonis and Martin T. Wells \thanks{Department of Statistics and Data Science, Cornell University; qas3@cornell.edu, mtw1@cornell.edu}
}
\date{Cornell University\\December 2024}

\begin{document}
\maketitle

\begin{abstract}
 This article explores the optimization of variational approximations for posterior covariances of Gaussian multiway arrays. To achieve this, we establish a natural differential geometric optimization framework on the space using the pullback of the affine-invariant metric. In the case of a truly separable covariance, we demonstrate a joint approximation in the multiway space outperforms a mean-field approximation in optimization efficiency and provides a superior approximation to an unstructured Inverse-Wishart posterior under the average Mahalanobis distance of the data while maintaining a multiway interpretation. We moreover establish efficient expressions for the Euclidean and Riemannian gradients in both cases of the joint and mean-field approximation. We end with an analysis of commodity trade data.
\end{abstract}
\noindent
\begin{small}
\end{small}

\section*{Introduction}
This article explores variational approximations to posterior distributions on covariances of Gaussian multiway data, where multiway data generally refers to data observations which themselves are multi-indexed, $\mathbf{Y}_{i} = \{(y_{i_{1}}, \ldots, y_{i_{k}}), i_{j} \in \{1,\ldots, d_{j}\}, j \in \{1,\ldots, D\} \}$. We can alternatively view this as observations coming in the form of $D$-way arrays, $\mathbf{Y}_{i} \in \mathbb{R}^{\times_{i = 1}^{D} d_{i}} $. In general, these data types come up in Gaussian data that exhibit some hierarchical factor structure, examples of which have been investigated in the context of global trade data \cite{hoff2011separable}, mortality data \cite{fosdick2014separable}, and conflict data \cite{hoff2011hierarchical}.

Analysis of multiway data often reduces to decomposing tensors using techniques such as the CP, Tucker, and Tensor Train \cite{oseledets2011tensor} decompositions. See \cite{kolda2009tensor} for a comprehensive review of these decompositions and their properties. MCMC is the gold standard for Bayesian inference on multiway data, with Gibbs techniques being developed for tensor regression \cite{guhaniyogi2017bayesian}, covariance inference under the Tucker product \cite{hoff2011separable}, and decompositions for missing data \cite{rai2014scalable}. Gibbs samplers, in practice, may yield an efficient \textit{per step} algorithm but are restricted to updating the full conditional posteriors. This poses two issues: updates are generally non-parallelizable, and in a setting where observations are tensor distributed, this often amounts to needing to compute expensive tensor reshaping arguments at each full conditional update step. Combining these two problems can usually create a computational bottleneck regarding Gibbs samplers' actual wall clock run time for multiway data.

Variational Bayesian techniques \cite{fox2012tutorial} are an approximate Bayesian computational method that has seen a growing interest in the last decade due to its computational procedure being an optimization problem rather than an MCMC sampling problem. Variational Bayes assumes that a posterior model may be sufficiently represented by an approximating family, and the goal then becomes finding the best member of that family. In other words, given a posterior model $\theta \vert Y$, one assumes a variational family on $\theta$, $Q_{\psi}(\theta)$, which is parameterized by the parameter set $\Psi$, and resolving the problem as finding the best such $\alpha \in \Psi$ such that $\theta \vert Y \approx Q_{\alpha}(\theta)$. Typically, this assumes the approximating distribution is an independently factorized distribution, $Q_{\alpha}(\theta) = \prod_{i = 1}^{D} Q_{\alpha_{i}}(\theta_{i})$, which is called the mean field approximation. Such an assumption for the approximating distribution is often not reasonable or, at the very least, questionable. Within the Bayesian literature, attempts at maintaining dependence structure in variational approximations have been explored \cite{petetin2021structured,wang2022structured}, and were shown to be a promising alternative to the mean field approximation.

A variational technique for nonparametric inference of multi-way data was developed in \cite{xu2013bayesian}. However, little work has been done for variational approximations in the parametric setting, the closest reference to our work being \cite{tran2021variational}, which explored using geometric optimization techniques as a natural methodology for the gradient descent of an ELBO defined on parameter spaces that admit a Riemann manifold structure, such as the sphere and the manifold of SPD matrices. Other geometry-related works include \cite{mccormack2023information}, which examined the Wishart information geometry of the 2-way case, and \cite{bouchard2021line}, who investigated a Riemannian optimization procedure for 2-way $t-$ distributed data.

With the goal of leveraging natural geometric structure within our optimization problem, we extend the existing literature to optimization of variational approximations for Inverse Wishart posteriors on the covariance of general $D$-way Gaussian data. Specifically, if $\mathbf{Y}_{i} \sim \mathcal{A}\mathcal{N}(\mathbf{0}, \{\Sigma_{i}\}_{i = 1}^{D})$, where $\{\Sigma_{i}\}_{i = 1}^{D}$ represent mode-wise covariances of an array normal likelihood, we examine variational approximations of posteriors defined on $\otimes_{i = D}^{1} \Sigma_{i}$,
which are the covariance structure arising from the vectorization of the array normal observations (see \cite{hoff2011separable} for further discussion of the tensor normal model). Our contributions in this paper may be summarized as:
\begin{itemize}
    \item We review the tensor normal model, alternative representations of the likelihood conducive to efficient gradient computations, and an efficient Monte Carlo sampler for an Inverse Wishart with a multiway scale matrix (Section \ref{sec: Mathematical Background}).
    \item We construct the pullback of the affine-invariant metric under the map $\phi: \times_{i = 1}^{D} \mathcal{P}(d_{i})$ $\rightarrow \mathcal{P}^{\otimes}(\prod_{i = 1}^{D} d_{i})$. This includes demonstrating the degeneracy of the naive pullback metric, and orthogonalization as a resolution to this degeneracy (Section \ref{sec: Geometry}).
    \item We construct the evidence lower bound (ELBO) of the joint approximation and mean field approximations, with corresponding Euclidean and Riemannian gradients, when applicable (Section \ref{sec: Variational Bayes}).
    \item We empirically demonstrate the superiority of the joint approximation in terms of iterations to convergence,  superiority of fit of the joint approximation over the mean field approximation with an unstructured prior baseline under average Mahalanobis distance. We then demonstrate the superiority of optimization of the joint model using the pullback metric over the product manifold metric and end by applying the joint variational approximation to the analysis of commodity trade data (Section \ref{sec: Empirical Comparisons}).
\end{itemize}

\section{Mathematical Background} \label{sec: Mathematical Background}
\subsection{Bayesian Inference}
Bayesian inference principally aims to infer a set of parameters $\theta$ having observed data parameterized by those parameters, $Y \sim L(\theta;Y)$ in the form of a posterior distribution through Bayes rule:
\begin{equation} \label{eq: Bayes rule}
    P(\theta \vert Y) = \frac{P(\theta;y) P(\theta)}{\int_{\theta} P(\theta;y) P(\theta) d\theta}.
\end{equation}
However, we obtain an analytic representation of \ref{eq: Bayes rule} is typically limited to only the simplest cases. Moreover, if such an analytic expression is tractible, efficient Monte Carlo inference is often limited by the dimension of the model, especially for constrained matrix-valued parameters.

In the cases where (\ref{eq: Bayes rule}) is not tractable, MCMC is often the gold standard for computational inference of a posterior distribution. Techniques for MCMC can be broadly categorized by Gibbs samplers, which requires iteratively updating full conditional distributions or random walk methods. Gibbs samplers have a clear disadvantage in that conditional updates are non-parallelizable. Hence, in circumstances where the parameter space itself is large in terms of the number of parameters, convergence is slow due to the need to iteratively update parameters, and autocorrelation accrued by the conditional updates. Random walk methods themselves are inefficient due to the necessity of small step sizes for updates to allow reasonable acceptance rates with the Metropolis ratio. A more advanced technique, Hamiltonian Monte Carlo (HMC), leverages an energy preserving differential equation to simulate dynamics on the unnormalized joint posterior. This allows first-order gradient information to be leveraged within parallelizable updates. While HMC is beneficial in regards to sampling from large parameter spaces, it necessarily requires multiple steps per iteration, and therefore requires several gradient-based steps per iteration.

Variational inference is a competitive alternative to any of these MCMC techniques, where instead we assume an approximate family of distributions parameterized by a collection of parameters $\alpha$, $Q_{\alpha}(\theta)$. The goal is instead assuming the family $Q$, find the $\alpha$ such that
\begin{equation}
    P(\theta \vert Y) \approx Q_{\alpha}(\theta)
\end{equation}
where the optimal $\alpha$ is found through 
\begin{equation} \label{eq: KL divergence}
\alpha = \arg \min_{\beta} KL[P(\theta \vert Y) || Q_{\beta}(\theta)].
\end{equation}
The expression in (\ref{eq: KL divergence}) can be decomposed into a more tractible form for optimization as
\begin{equation} \label{eq: ELBO}
    P(Y) = \mathbb{E}_{Q_{\alpha}}[\log P(X,\theta) - Q_{\alpha}(\theta)] - KL[P\vert \vert Q_{\alpha}].
\end{equation}
Noting that $P(Y)$ is fixed with respect to $\alpha$, maximizing $\mathbb{E}_{Q_{\alpha}}[\log P(X,\theta) - Q_{\alpha}(\theta)]$ through $\alpha$ implicitly  minimizes $KL[P\vert \vert Q_{\alpha}]$.

\subsection{Tensor Normal}
The tensor observations $\mathcal{Y}^{i} \sim TN(\mathbf{M}, \Sigma_{d} \circ \Sigma_{d-1} \circ \cdots \circ \Sigma_{2} \circ \Sigma_{1})$, where $TN(a,b)$ denotes the tensor normal model with mean $a$ and covariance array $b$. Then for $k \in \{1,\ldots, n\}$, let $\mathcal{Y}^{i}_{(k)}$ denote the mode $k$ matricization. Letting $MN(a,B,C)$ denote the matrix normal model with mean $a$, row covariance $B$ and column covariance $C$ (see \cite{dutilleul1999mle} for a detailed description of the matrix normal model).  The tensor normal can be characterized by vectorization operations.
\begin{proposition}
    Let $\mathcal{Y} \sim TN(\mathbf{M}, \Sigma_{d} \circ \Sigma_{d-1} \circ \cdots \circ \Sigma_{2} \circ \Sigma_{1})$, then
    \begin{align}
    \mathcal{Y}_{(d)} &\sim MN(\mathbf{M}_{(d)}, \Sigma_{d}, \Sigma_{1} \otimes \Sigma_{2} \otimes \cdots \otimes \Sigma_{d-1}) \\
    vec(Y_{(d)}) &\sim N(vec(M_{(d)}), \Sigma_{1} \otimes \Sigma_{2} \otimes \cdots \otimes \Sigma_{d}).
    \end{align}
    \begin{proof}
        By \cite{hoff2011separable}, the density of the tensor normal is given as
        \[
p(Y|M, \Sigma_1, \ldots, \Sigma_K) = (2\pi)^{-d/2} \left( \prod_{k=1}^{K} |\Sigma_k|^{-d/(2d_k)} \right) \exp\left(-\frac{1}{2} \| (\mathcal{Y}-{\bf M}) \times \Sigma^{-1/2} \|^{2} \right),
\]
We can consider the distribution of $Y$ as the Tucker product between the list $\Xi = \{\Sigma_{d}^{1/2}, \ldots, \Sigma_{1}^{1/2} \}$ and a standard tensor normal, $Z \sim TN({\bf M}, I_{d} \circ I_{d-1} \circ \cdots \circ I_{1})$
\begin{align}
    \mathcal{Y} &= Z \times \Xi + {\bf M} = Z \times_{1} \Sigma_{d}^{1/2} \times_{2} \Sigma_{d - 1}^{1/2} \circ \cdots \circ \times_{d} \Sigma_{1}^{1/2} + \bf{M} \\
    \implies \mathcal{Y}_{(d)} &= \Sigma_{d}^{1/2} Z_{(d)} (\Sigma_{1}^{1/2} \otimes \Sigma_{2}^{1/2} \otimes \cdots \otimes \Sigma_{d - 1}^{1/2}) \nonumber \\
    & \quad \quad + \; {\bf M_{(d)}} \sim MN({\bf M_{(d)}}, \Sigma_{d}, \Sigma_{1} \otimes \Sigma_{2} \otimes \cdots \otimes \Sigma_{d-1}) \\
    \implies vec(\mathcal{Y}_{(d)}) &\sim N(vec({\bf M}), \Sigma_{1} \otimes \Sigma_{2} \otimes \cdots \otimes \Sigma_{d - 1} \otimes \Sigma_{d}).
\end{align}
\end{proof}
\end{proposition}

\subsection{Tensors, Matricization, and Multilinear to Linear Mappings} \label{sec: Tensor}
In this section we review various tensor representation results and give some useful novel identities.
\begin{definition}
    For $A \in \mathbb{R}^{m \times n}$, $B \in \mathbb{R}^{p \times q}$ the Kronecker product, $A \otimes B \in \mathbb{R}^{mp \times nq}$ is defined:
    \[
    A \otimes B = \begin{pmatrix}
        a_{11} B & a_{12} B & \cdots &a_{1m} B \\
        \vdots & \vdots & \ddots & \vdots \\
        a_{n1} B & a_{n2} B & \cdots & a_{nm}B.
    \end{pmatrix}
    \]
\end{definition}
Note $(A\otimes B)_{(r - 1)p + v, (s - 1)q + w} = a_{rs} b_{vw}$. This is expressed more generally in the following theorem.
\begin{theorem} \label{thm: general kronecker indices}
Let $A^{(k)} \in \mathbb{R}^{q_{k} \times p_{k}}$, and let $i_{1}^{(k)}$, $i_{2}^{(k)}$ be such that $1 \leq i_{1}^{(k)} \leq q_{k}$, $1 \leq i_{2}^{(k)} \leq p_{k}$, then
    \[
    (\otimes_{i = 1}^{D} A^{(k)})_{\sum_{i = 1}^{D-1} (i_{1}^{(i)} - 1) \prod_{k > i} q_{k} + i_{1}^{(D)}, \sum_{j = 1}^{D-1} (i_{2}^{(j)} - 1) \prod_{k > j} p_{k} + i_{2}^{(D)}} = \prod_{k = 1}^{D}A^{(k)}_{i_{1}^{(k)}, i_{2}^{(k)}}.
    \]
\end{theorem}

Let $S \in \mathbb{R}^{\times_{i = 1}^{D} d_{i}}$ be a tensor array with $D$ modes, $M = \{1,\ldots, D\}$. Per \cite{kolda2009tensor}, the mode $k$ matricization of $S$, $S_{(k)} \in \mathbb{R}^{d_{k} \times \prod_{i \in M/k} d_{i}}$, $k \in \{1,\ldots, D\}$, maps the tensor element $(i_{1}, \ldots, i_{D})$ to the matrix element $(i_{k}, j)$ as:
\[
j = 1 + \sum_{k = 1 \not = n}^{D}(i_{k} - 1)J_{k} \quad J_{k} = \prod_{m = 1 \not= n}^{k - 1} d_{m}.
\]
In this way, we can analogously view multilinear to linear mappings of indices. Suppose $S \in \mathcal{P}(d)$ such that $d = \prod_{i = 1}^{k} d_{i}$. 

\begin{definition}
    Let $\mathcal{S}_{D}(k)$ be the set of $2D$ way tensors that are symmetric with respect to modes $k$ and $2k$. That is, $\mathcal{T} \in \mathcal{S}_{D}(k)$ if
    \[
    \mathcal{T}[i_{1}^{(1)}, \ldots,i_{1}^{(k)}, \ldots, i_{1}^{(D)}, i_{2}^{(1)}, \ldots, i_{2}^{(k)}, \ldots, i_{2}^{(D)}] = \mathcal{T}[i_{2}^{(1)},\ldots,i_{2}^{(k)},\ldots, i_{1}^{(D)},  i_{2}^{(1)}, \ldots,i_{1}^{(k)}, \ldots, i_{1}^{(D)}].
    \]
    Moreover, for the set $M = \{ i_{a_{1}}, \ldots, i_{a_{k}}\}$, let $\mathcal{S}_{D}(M)$ be the set of $2D$ way tensors such that $\mathcal{T} \in \mathcal{S}_{D}(q)$ for all $q \in M$.
\end{definition}

\begin{definition} \label{def:Symmetric Folding}
    Let $A \in \mathbb{S}(\prod_{i = 1}^{D} d_{i})$, define the \textbf{symmetric folding} of $\mathcal{F}: A \rightarrow \mathcal{A} \in \mathbb{R}^{d_{1} \times d_{2} \times \cdots \times d_{D} \times d_{1} \times \cdots \times d_{D}}$ as
    \[
    A[p,q] \rightarrow \mathcal{A}[i_{1}^{(1)}, i_{1}^{(2)}, \ldots, i_{1}^{(D)}, i_{2}^{(1)}, \ldots, i_{2}^{(D)}]
    \]
    where $p = \sum_{k = 1}^{d - 1}(i_{1}^{(k)} - 1)\prod_{j = k+1}^{D - 1} d_{j} + i_{1}^{D}, \quad q = \sum_{k = 1}^{d - 1}(i_{2}^{(k)} - 1)\prod_{j =  k+ 1}^{D - 1}d_{j} + i_{2}^{D}$.
\end{definition}

Note that by Theorem \ref{thm: general kronecker indices}, it immediately follows that $\mathcal{F}(\otimes_{i = 1}^{D} A_{i}) \in \mathcal{S}_{D}(\{1,\ldots, D\})$ if $A_{i} \in \mathbb{S}(d_{i})$.

\begin{definition} \label{def: Symmetric Unfolding}
     Let $\mathcal{A} \in \mathcal{S}_{D}(\{1,\ldots, D\})$ and $T = \{1,\ldots, D\}$. Define the \textbf{symmetric unfolding} of $ \mathcal{U}\mathcal{F}: \mathcal{A} \rightarrow \mathcal{A}_{(T)} \in \mathcal{S}(\prod_{i \in T} d_{i})$ as
    \[
    A[p,q] = \mathcal{A}[i_{1}^{(1)}, i_{1}^{(2)}, \ldots, i_{1}^{(D)}, i_{2}^{(1)}, \ldots, i_{2}^{(D)}],
    \]
    where $p,q$ were defined in Definition \ref{def:Symmetric Folding}.
\end{definition}

Note that Definitions \ref{def:Symmetric Folding} and \ref{def: Symmetric Unfolding} is generally expression for the mode folding of a matrix and the mode unfolding of a tensor (see \cite{kolda2009tensor} for further discussion).

\begin{proposition} \label{prop: Elementary Components}
    Suppose $A \in \mathbb{R}^{\prod_{i = 1}^{D} d_{i} \times \prod_{i = 1}^{D} d_{i}}$ and let $E_{k}^{a, b}$ be the matrix such that
    \[
    E_{k}^{a, b}[m,n] = \begin{cases}
        1 \text{ if }  [m,n] = [a, b] \\
        0 \text{ otherwise }.
    \end{cases}
    \]
    Then $A$ can be written as
    \[
    A = \sum_{i_{1}^{(1)},i_{2}^{(1)} \in [ d_{1}]} \cdots \sum_{i_{1}^{(D)},i_{2}^{(D)} \in [ d_{D} ]} A_{i,j} \otimes_{k = 1}^{D} E_{k}^{i_{1}^{(k)}, i_{2}^{(k)}}
    \]
    where $i = \sum_{k = 1}^{D - 1} (i_{1}^{(k)} - 1) \prod_{n = k +1}^{d} d_{n} + i_{1}^{(D)}$, $j = \sum_{k = 1}^{D - 1} (i_{2}^{(k)} - 1) \prod_{n = k +1}^{D} d_{n} + i_{2}^{(D)}$, and $[p] = \{1,\ldots, p\}$ for $p \in \mathbb{N}$.
    \begin{proof}
        This can be straightforwardly observed by letting $\mathcal{E}^{[i_{1}^{(1)}, i_{2}^{(1)}, \ldots, i_{1}^{(D)}, i_{2}^{(D)}]}$ be the tensor such that
        \[
        \mathcal{E}^{[i_{1}^{(1)}, i_{2}^{(1)}, \ldots, i_{1}^{(D)}, i_{2}^{(D)}]}[p_{1}, p_{2}, \ldots, p_{2D}] = 
        \begin{cases}
            1 \text{ if } [p_{1}, p_{2}, \ldots, p_{2D}] = [i_{1}^{(1)}, i_{1}^{(2)}, \ldots, i_{1}^{(D)}, i_{2}^{(1)}, \ldots, i_{2}^{(D)}] \\
            0 \text{ otherwise}.
        \end{cases}
        \]
        Let $\mathcal{F}(A) = \mathcal{A} \in \mathbb{R}^{(\times_{i = 1}^{D} d_{i}) \times (\times_{i = 1}^{D} d_{i})}$, and note that
        \begin{align*}
        \mathcal{A} &= \sum_{i^{(1)}_{1},i^{(1)}_{1} \in [d_{1}]} \cdots \sum_{i^{(D)}_{1},i^{(D)}_{2} \in [d_{D}]} \mathcal{A}[i_{1}^{(1)}, \ldots, i_{2}^{(D)}] \odot \mathcal{E}^{[i_{1}^{(1)}, \ldots,  i_{2}^{(D)}]}
        \end{align*}
        where '$\odot$' denotes the element wise product of two tensors. Per Definition \ref{def: Symmetric Unfolding},
        \begin{align*}
             \mathcal{E}_{(1,\ldots,D)}[i,j]  = \begin{cases}
                 1 \text{ if } i =  \sum_{k = 1}^{D - 1}(i_{1}^{(k)} - 1)\prod_{j = k+1}^{D - 1} d_{j} + i_{1}^{(D)}, \quad j = \sum_{k = 1}^{d - 1}(i_{2}^{(k)} - 1)\prod_{j =  k+ 1}^{D - 1}d_{j} + i_{2}^{(D)} \\
                 0 \text{ otherwise.}
             \end{cases}
         \end{align*} 
         Hence by Theorem \ref{thm: general kronecker indices}, \[
         \mathcal{E}_{(1,\ldots,D)} = \otimes_{k = 1}^{D} E_{k}^{i_{1}^{(k)}, i_{2}^{(k)}}
        \]
        where
        \[
        E_{k}^{i_{1}^{(k)}, i_{2}^{(k)}}[m,n] = \begin{cases}
            1 \text{ if }  [m,n] = [i_{1}^{(k)}, i_{2}^{(k)}] \\
            0 \text{ otherwise. }
        \end{cases}
        \]
        By linearity of the operator in Definition \ref{def: Symmetric Unfolding} and noting $\mathcal{A}_{\{1,\ldots, D\}} = A$, the result follows immediately.
    \end{proof}
\end{proposition}
We will make use of Proposition \ref{prop: Elementary Components} to construct a tensor normal model which is conducive to efficient gradient computations, as described in the following proposition:
\begin{proposition} \label{prop: efficient trace}
    Suppose $\mathcal{Y}_{1}, \ldots, \mathcal{Y}_{n} \sim \mathcal{T}\mathcal{N}(\mathbf{0}, \Sigma_{D} \circ \cdots \circ \Sigma_{1})$, letting $y_{i} = vec(\mathcal{Y}_{i}) \in \mathbb{R}^{\prod_{i = 1}^{D} d_{i}}$ and $S = \sum_{i = 1}^{n} y_{i} y_{i}^{T}$ then
    \begin{equation} \label{eq: kronecker trace}
        tr([\otimes_{i = 1}^{D} \Sigma_{i}]^{-1} S) =  tr(\Sigma_{i}^{-1} T^{i}(S, \Sigma_{-i})),
    \end{equation}
    where
    \begin{align*}
        T^{(k)}(S,\Sigma_{-i}) &= \frac{C^{k}(S, \Sigma_{-i})) + [C^{k}(S, \Sigma_{-i}))]^{T}}{2}\\
        C^{k}(S, \Sigma_{-i}))[i^{(k)}_{1}, i^{(k)}_{2}] &=  \sum_{\substack{i_{1}^{(t)} \leq i_{2}^{(t)} \\ t \neq k}} \prod_{ t \in \{1,\ldots, D\} / k} A_{t}^{-1}[i_{1}^{(t)}, i_{2}^{(t)}] S_{\Gamma(i^{(k)}_{1},i^{(k)}_{2})}[i, j]E_{k}^{i_{1}^{(k)}, i_{2}^{(k)}}, \quad i_{1}^{(k)} \leq i_{2}^{(k)} 
    \end{align*}
    such that $i = \sum_{k = 1}^{d - 1} (i_{1}^{(k)} - 1) \prod_{n = k +1}^{d} d_{n} + i_{1}^{(d)}$, $j = \sum_{k = 1}^{d - 1} (i_{2}^{(k)} - 1) \prod_{n = k +1}^{d} d_{n} + i_{2}^{(d)}$ and $S_{\Gamma(i^{(k)}_{1},i^{(k)}_{2})} = \mathcal{U}\mathcal{F}(\mathcal{S}_{\Gamma[i^{(k)}_{1},i^{(k)}_{2}]})$ where $\mathcal{S}_{\Gamma[q]}[\cdot] = \sum_{J \subset I: i_{J}^{(1)} \neq i_{J}^{(2)}, q \not \in J} \mathcal{F}({S})[\cdot]$ with $I = \{1,\ldots, D\}$ being the mode indices.
    \begin{proof}
        First note that by Proposition \ref{prop: Elementary Components}, 
        \begin{align*}
        tr([\otimes_{i = 1}^{D} \Sigma_{i}]^{-1} S) &= \sum_{i_{1}^{(1)},i_{2}^{(1)} \in [d_{1}]} \cdots \sum_{i_{1}^{(d)},i_{2}^{(d)} \in [d_{D}]} tr([\otimes_{i = 1}^{D} \Sigma_{i}]^{-1} S_{i,j} \otimes_{k = 1}^{D} E_{k}^{i_{1}^{(k)}, i_{2}^{(k)}}) \\
        &= \sum_{i_{1}^{(1)},i_{2}^{(1)} \in [d_{1}]} \cdots \sum_{i_{1}^{(D)},i_{2}^{(D)} \in [d_{D}]} tr([ S[i,j] \big[\otimes_{k = 1}^{D} \Sigma_{i}^{-1}E_{k}^{i_{1}^{(k)}, i_{2}^{(k)}} \big])\\
        &= \sum_{i_{1}^{(1)},i_{2}^{(1)} \in [d_{1}]} \cdots \sum_{i_{1}^{(D)},i_{2}^{(D)} \in [d_{D}]} S[i,j] \prod_{k = 1}^{D} \Sigma_{k}^{-1}[i_{1}^{(k)}, i_{2}^{(k)}].
        \end{align*}
        Now observe 
        \begin{align*}
        &\sum_{i_{1}^{(1)},i_{2}^{(1)} \in [d_{1}]} \cdots \sum_{i_{1}^{(D)},i_{2}^{(D)} \in [d_{D}]} S[i,j] \prod_{k = 1}^{D} \Sigma_{k}^{-1}[i_{1}^{(k)}, i_{2}^{(k)}] \\
        &= tr\big(\Sigma_{q}^{-1} \sum_{i_{1}^{(1)},i_{2}^{(1)} \in [d_{1}]} \cdots \sum_{i_{1}^{(D)},i_{2}^{(D)} \in [d_{D}]} \big(S[i,j] \prod_{k \neq q}^{D} \Sigma_{k}^{-1}[i_{1}^{(k)}, i_{2}^{(k)}] \big) E_{q}^{i_{1}^{(q)},i_{2}^{(q)}} \big).
        \end{align*}
        And by folding $S$ into the tensor $\mathcal{F}(S) = \mathcal{S} \in \mathbb{R}^{(\times_{i = 1}^{D} d_{i}) \times (\times_{i = 1}^{D} d_{i})}$,
        \begin{align}
            &\sum_{i_{1}^{(1)},i_{2}^{(1)} \in [d_{1}]} \cdots \sum_{i_{1}^{(D)},i_{2}^{(D)} \in [d_{D}]} \big(S[i,j] \prod_{k \neq q}^{D} \Sigma_{k}^{-1}[i_{1}^{(k)}, i_{2}^{(k)}] \big) E_{q}^{i_{1}^{(q)},i_{2}^{(q)}} \label{eq: entries of trace} \nonumber \\
            \nonumber &= \sum_{i_{1}^{(1)},i_{2}^{(1)} \in [d_{1}]} \cdots \sum_{i_{1}^{(D)},i_{2}^{(D)} \in [d_{D}]} \big(\mathcal{S}[i_{1}^{(1)}, i_{1}^{(2)}, \cdots,i_{1}^{(D)}, i_{2}^{(1)}, \ldots, i_{2}^{(D)} ] \prod_{k \neq q}^{D} \Sigma_{k}^{-1}[i_{1}^{(k)}, i_{2}^{(k)}] \big) E_{q}^{i_{1}^{(q)},i_{2}^{(q)}}. \nonumber \\
            \end{align}
        Since $\mathcal{S} \not \in \mathcal{S}_{D}(T)$ for any $T \subset \{1,\ldots, D\}$, the upper triangular entries of \ref{eq: entries of trace} may be written using only the upper triangular entries of $\{\Sigma_{k}\}_{k = 1}^{D}$ as
        \begin{equation} \label{eq: upper triangular entries of trace}
            \sum_{i_{1}^{(1)} \leq i_{2}^{(1)} \in [k_{1}]} \cdots \sum_{i_{1}^{(d)} \leq i_{2}^{(d)} \in [k_{d}]} \big(\mathcal{S}_{\Gamma[q]}[i_{1}^{(1)}, i_{1}^{(2)}, \cdots,i_{1}^{(D)}, i_{2}^{(1)}, \ldots, i_{2}^{(D)} ] \prod_{k \neq q}^{D} \Sigma_{k}^{-1}[i_{1}^{(k)}, i_{2}^{(k)}] \big) E^{i_{1}^{(q)},i_{2}^{(q)}},
        \end{equation}
        where $\mathcal{S}_{\Gamma[q]}[\cdot] = \sum_{J \subset I: i_{J}^{(1)} \neq i_{J}^{(2)}, q \not \in J} \mathcal{S}[\cdot]$. Letting $S_{\Gamma[q]} = \mathcal{U}\mathcal{F}(\mathcal{S}_{\Gamma[q]})$, and noting (\ref{eq: entries of trace}) is symmetric, the proof is completed by symmetrizing (\ref{eq: upper triangular entries of trace}).
    \end{proof}
\end{proposition}
To interpret Proposition \ref{prop: efficient trace}, note that each of $\{\Sigma_{j}\}_{j = 1}^{D}$ and $S$ are symmetric, however $S$ is not symmetric in regards to interchanging indices of $\{\Sigma_{j} [i_{1}^{(j)}, i_{2}^{(j)}]\}_{j = 1}^{D}$. Efficient computation of the term \ref{eq: kronecker trace} boils down to leveraging the symmetry of all the components involved. To effectively do this, Proposition \ref{prop: efficient trace} states that we have a pre-computation step involving computing the sum of the permutation matrices in the construction of $S_{\Gamma[q]}$ for $q \in \{1,\ldots, D\}$. 

To our knowledge, the only alternative to computing the gradient of equation (\ref{eq: kronecker trace}) would be through matricization of the original tensor in mode instead of vectorization. Letting $\mathcal{Y}_{(k)}$ denote the matricization of a tensor observation with $\mathcal{Y} \sim \mathcal{T}\mathcal{N}(\mathbf{0}, \Sigma_{D} \circ \cdots \circ \Sigma_{1})$, then
\begin{equation} \label{eq: trace matricization}
    tr(\| \mathcal{Y} \times \Sigma^{-\frac{1}{2}} \|^{2}) = tr(\Sigma_{k}^{-1} \mathcal{Y}_{(k)} (\otimes_{j \neq k}^{D} \Sigma_{j}^{-1}) \mathcal{Y}_{(k)}^{T}).
\end{equation}
With $\mathcal{Y}_{(k)} \in \mathbb{R}^{d_{k} \times \prod_{j \neq k}^{D} d_{j}}$, $\otimes_{j \neq k}^{D} \Sigma_{j}^{-1} \in \mathcal{P}(\prod_{j \neq k}^{D} d_{j})$, computation of $\mathcal{Y}_{(k)} (\otimes_{j \neq k}^{D} \Sigma_{j}^{-1}) \mathcal{Y}_{(k)}^{T}$ is necessarily at least $\mathcal{O}( [d_{k} \prod_{j \neq k}^{D} d_{j}^{2} + d_{k}^{2} \prod_{j \neq k}^{D} d_{j}])$. Moreover, this complexity scales multiplicatively with the number of number of samples is necessary as no trace trick can be leveraged like in the vectorized case. Compared to equation (\ref{eq: upper triangular entries of trace}) requires $D-1$ total multiplications summed up over a total of $\prod_{i \neq k}^{D} \frac{d_{i}(d_{i} + 1)}{2}$ combined entries in $\{\Sigma_{i}\}_{i \neq k}^{D}$. This yields a complexity of $\approx \mathcal{O}(\frac{D-1}{2^{D - 1}} \prod_{i \neq k}^{D} d_{i}^{2})$ in the computation of $T^{k}(S, \Sigma_{-k})$ in (\ref{eq: kronecker trace}), and does not increase the scale with the number of samples. Note that these computations can be done in parallel, so the complexity of (\ref{eq: kronecker trace}) will be bottlenecked by $\mathcal{O}(\max_{k}[\frac{D-1}{2^{D - 1}} \prod_{i \neq k}^{D} d_{i}^{2}])$.

\subsection{Monte Carlo Simulation for an Inverse Wishart with Multiway Scale Parameter}
Suppose $\Sigma \sim W_{d}(\eta, Q)$, typical simulation of an unstructured Wishart is to generate $X \in \mathbb{R}^{d \times \eta}$ with $X_{i,j} \sim N(0,1)$, compute $L = \mathcal{L}(Q)$, where $\mathcal{L}(\cdot)$ is the Cholesky factor of a matrix, then:
\begin{equation} \label{eq: Wishart Sampling}
    \Sigma \sim LX (LX)^{T}.
\end{equation}

Naively, such simulation is $\mathcal{O}(d^{3})$. Hence, computing Monte Carlo estimates of the distribution of matrix statistics such as $tr(\Sigma)$ or $\log \vert \Sigma \vert$ is a challenge when $d$ is large. However, in the setting where $Q = \otimes_{i = 1}^{D} Q_{i}$, we can utilize properties of tensor operations to quickly draw Monte Carlo samples.

First note when $Q = \otimes_{i = 1}^{D} Q_{i}$, $\mathcal{L}(\otimes_{i = 1}^{D} Q_{i}) = \otimes_{i = 1}^{D} \mathcal{L}(Q_{i})$, then we can express
\[
\Sigma \sim (\otimes_{i = 1}^{D}\mathcal{L}(Q_{i})) X ((\otimes_{i = 1}^{D}\mathcal{L}(Q_{i})) X)^{T}.
\] 
Now consider the generation of the random tensor $\mathcal{X} \in \mathbb{R}^{(\times_{i = D}^{1} d_{i}) \times \eta}$ such that $\mathcal{X}_{\cdot} \sim N(0,1)$ elementwise. Note that the order of dimensions here is reversed to that of the Kronecker product. Letting $L_{i} = \mathcal{L}(Q_{i})$, Observe that
\begin{align}
    \mathcal{Y} &= (L_{1}, \ldots, L_{D})^{\frac{1}{2}} \circ \mathcal{X} \\
    \implies &\mathcal{UF}(\mathcal{Y}| 1,\ldots, D) = (\otimes_{i = 1}^{D} L_{i}) X.
\end{align}
However, the main burden of simulating from a Wishart with a Kronecker structured scale matrix then moves the burden of constructing $\otimes_{i = 1}^{D} L_{i}$ to the computation of $XX^{T}$. Note that for this task, we can instead observe that the Bartlett decomposition \cite{kshirsagar1959bartlett} provides an efficient technique for such a decomposition. Specifically, let $L$ be the lower triangular matrix such that
\begin{align*}
    L_{i,i} \sim \sqrt{\chi_{\nu - i + 1}^{2}} \quad \text{and} \quad L_{i,j} \sim \mathcal{N}(0,1) \text{ for } i >j.
\end{align*}
Then $LL^{T} \sim \mathcal{W}_{D}(\nu, I)$. Note that under the Bartlett decomposition, the simulation of $L$ can be performed in $\approx \mathcal{O}(\frac{1}{2} D^{3})$ rather than $\mathcal{O}(\nu D^{3})$. As $\nu$ is necessarily on the order of $D$, this can substantially reduce the computation time when $D$ is large, as will be the case in most of our examples. Reshaping $L \in \mathbb{R}^{\prod_{i = 1}^{D} d_{i}^{2}}$ into the $2D$ array $L \rightarrow \mathcal{L} \in \mathbb{R}^{d_{D} \times d_{D - 1} \times \cdots d_{1} \times d_{D} \times \cdots d_{1}}$, then note
\begin{align}
    \mathcal{G} = (L_{D}, \ldots, L_{1}) \circ \mathcal{L}  \implies  \mathcal{UF}(\mathcal{G}| 1,\ldots, D) = (\otimes_{i = 1}^{D} L_{i}) L.
\end{align}
Where $(L_{D}, \ldots, L_{1}) \circ \Omega$ denotes the sequential operation: for $i \in \{1,\ldots, D\}$, $\Omega \leftarrow L_{i} \times_{i} \Omega$ where $\times_{i}$ is the mode-$i$ Tucker product $L_{i} \times_{i} \Omega = L_{i} \Omega_{(i)}$.
The point is that this representation reduces the large matrix multiplication of $(\otimes_{i = 1}^{D} L_{i}) L$ to smaller sequential mode-wise operations. The complexity of a $A \times_{i} \mathcal{B}$ for $\mathcal{B} \in \mathbb{R}^{\times_{i = 1}^{D} p_{i}}$ is in general $\mathcal{O}(p_{i}^{3} \prod_{j \neq i} p_{j})$ when $A$ is square. Hence, the naive matrix multiplication being $\approx \mathcal{O}(\frac{1}{2} \prod_{i = 1}^{D} d_{i}^{3})$ then simplifies to $\approx \mathcal{O}(\sum_{i = 1}^{D} d_{i}^{3} \prod_{j \neq i} d_{j}^{2})$.

In general, then, Cholesky factors of Inverse-Wishart (or Wishart with an inverted scale matrix) can be computed as in Algorithm 1.

\begin{algorithm}[H]
\caption{Multi-way Cholesky Inverse-Wishart Monte Carlo}
\begin{algorithmic}[1]  % [1] for line numbering
    \STATE Initialize with $\{L_{1}, \ldots, L_{D}\}$, $\nu_{v}$, and $L$
    \STATE Reshape $L \rightarrow \mathcal{L} \in \mathbb{R}^{d_{D} \times d_{D - 1} \times \cdots \times d_{1} \times d_{D} \times \cdots \times d_{1}}$
    \FOR{$i \in \{1, \ldots, D\}$}
        \STATE Compute $\mathcal{L} \leftarrow L_{D - i + 1} \times_{i} \mathcal{L}$
    \ENDFOR
    \STATE Let $W_{L} \leftarrow \mathcal{L}_{(1,\ldots, D)}$
    \STATE Return $W_{L}$
\end{algorithmic}
\end{algorithm}

As a comparison, the mean field approximation is straightforward to simulate from, where cholesky factors are independently simulated according to $L_{i} \stackrel{d}{=} \mathbb{L}(A_{i}) L_{i}$ with 
\[
L_{i}[m,n] = \begin{cases}
    \sqrt{\chi^{2}_{\nu_{i} - d_{i} + 1}} \text{ if } m = n\\
    N(0,1) \text{ if } m > n.
\end{cases}
\] 
When samples are taken under the mean field approximation, then simulated by simply taking their Kronecker product.

\section{The Geometry of Multiway SPD Matrices Under the Affine-Invariant Metric} \label{sec: Geometry}
\subsection{General Differential Geometry}
\begin{definition}
    \textbf{Topological manifold:} A topological manifold, $\mathcal{M}$, is a second countable Hausdorff space such that for every $q \in \mathcal{M}$, there exists a neighborhood containing $q$ that is homeomorphic to the Euclidean space.
\end{definition}
\begin{definition}
    \textbf{Tangent Space:} Let $q \in \mathcal{M}$ be a point on a manifold $\mathcal{M}$. The tangent space $T_{q}\mathcal{M}$ at $q$ on $\mathcal{M}$ is defined as the equivalence class of time derivatives of curves:
    \[
    T_{q}\mathcal{M} := \{\gamma'(0): \gamma(0) = q, \gamma(t) \in \mathcal{M} \quad t \in \mathbb{R}^{\geq 0}\}
    \]
\end{definition}
\begin{definition}
    \textbf{Riemannian Manifold:} A Riemannian manifold is a pair, $(\mathcal{M}, g)$, where $\mathcal{M}$ is a topological manifold, and $g$ is a bilinear positive definite inner product on the tangent space called a Riemannian metric.
\end{definition}

In this work, we consider the affine-invariant metric as the choice of the Riemannian metric. This is defined for $U,V \in T_{\Sigma} \mathcal{P}(d)$ as
\begin{equation} \label{eq: Affine Invariant metric}
    g_{\Sigma}^{AI}(U,V) = tr(\Sigma^{-1} U \Sigma^{-1}V).
\end{equation}
Although other metrics exist, such as the Log-Cholesky, Bures-Wasserstein, and log-Euclidean (see \cite{lin2019riemannian}, \cite{huang2015log} and \cite{han2021riemannian} for descriptions of these metrics), the Affine-Invariant is a natural choice in the optimization of covariances of a multivariate normal due to it's connection with the Fisher information.
\begin{definition}
    \textbf{Geodesic:} A geodesic is the notion of a straight line on a manifold. For a Riemannian manifold $(\mathcal{M},g)$, it is a path $\gamma(t) \in \mathcal{M}$ such that $\frac{d}{dt}\| \gamma(t)\|_{G} = 0$. The geodesic that connects the points $\gamma(a), \gamma(b) \in \mathcal{M}$ is constructed as the path, $\gamma$ which minimizes
    \begin{equation} \label{eq: geodesic problem}
    E(\gamma) = \frac{1}{2} \int_{a}^{b} \|\dot{\gamma}(t)\|_{g} dt.
    \end{equation}
\end{definition}
Given an initial point $\Sigma(0) \in \mathcal{P}(d)$ and an initial tangent vector $V(0) \in \mathcal{S}(d)$, a simple closed-form solution to equation (\ref{eq: geodesic problem}) under the affine-invariant metric for general SPD matrices was derived in \cite{moakher2006symmetric} as
    \begin{equation}
        \Sigma(t) = \Sigma(0)^{\frac{1}{2}} \exp (t \Sigma(0)^{-1/2} V(0) \Sigma(0)^{-\frac{1}{2}}) \Sigma(0)^{\frac{1}{2}}.
    \end{equation}
\begin{definition} \label{def: Geodesic convexity}
    \textbf{Geodesic Convexity} Let $(\mathcal{M},g)$ be a Riemannian manifold and $f$ be a smooth function defined on $ A \subseteq \mathcal{M}$. Then $f$ is said to be geodesically convex on $A$ if for any $p,q \in A$ and $\gamma(t)$ is the unique geodesic connecting $p$ and $q$ such that $\gamma(0) = p$, $\gamma(1) = q$,
    \[
    f(\gamma(t)) \leq tq + (1-t)p.
    \]
\end{definition}

\begin{definition} \label{def: Conditional Geodesic Convexity}
    \textbf{Conditional Geodesic Convexity} Let $(\times_{i = 1}^{d} \mathcal{M}_{i}, \oplus_{i = 1}^{d} G_{i})$ be a product of Riemannian manifolds and $f$ be a smooth function defined $ \times_{i = 1}^{d}A_{i} \subseteq \times_{i = 1}^{d} \mathcal{M}_{i}$. Then $f$ is said to be conditionally geodesically convex on $A$ if for any $\{p_{i},q_{i}\}_{i = 1}^{d} \in \times_{i = 1}^{d}A_{i}$ and $\gamma_{i}(t)$ is the unique geodesic connecting $p_{i}$ and $q_{i}$ such that $\gamma_{i}(0) = p_{i}$, $\gamma(1) = q_{i}, f_{i}(\gamma_{i}(t)\vert \{p_{-i}, q_{-i}\}) \leq tq_{i} + (1-t)tp_{i}.$  That is, conditional on $\{p_{-i}, q_{-i}\}$, $f$ is geodesically convex on $A_{i}$.
\end{definition}

It was shown in \cite{wiesel2012geodesic} that the space of Kronecker structured spd matrices is geodesically convex for the covariance of a multivariate normal. Given that the Inverse Wishart is a conjugate family to the multivariate normal, all ELBO computations considered later in this paper are geodesically convex. As will be shown in the next section, we consider the geometry of the space of Kronecker structured matrices by pulling back the geometry from the Kronecker space to the Cartesian product, wherein geometric computations are defined to only the components, where it is well known that each component manifold is geodesically complete (see \cite{moakher2011riemannian} for further discussion). Hence, leveraging the geometry of this space allows for a well-defined, geometrically aware convex optimization problem. 

\subsection{Pullback Geometry of d-Kronecker SPD Matrices}\label{sec: SPD matrices}
A covariance matrix $\Sigma \in \mathcal{P}(\prod_{i = 1}^{d} n_{i})$ is d-Kronecker if it can be decomposed as
\[
\Sigma = \otimes_{i = 1}^{d} \Sigma_{i}
\]
where $\Sigma_{i} \in \mathcal{P}(n_{i})$. The differential of a separable covariance matrix follows a product rule:
\[
d(\otimes_{i = 1}^{d} \Sigma_{i}) = \sum_{i = 1}^{d} [\otimes_{j = 1}^{d} (1_{j = i} d\Sigma_{i} + 1_{j \neq i} \Sigma_{i})].
\]
Hence, the tangent space is expressible as
\[
\mathcal{T}_{\otimes_{i = 1}^{d} \Sigma_{i}} \times_{i = 1}^{d} \mathcal{P}(n_{i}) =  \{ \sum_{i = 1}^{d} [\otimes_{j = 1}^{d} (1_{j = i} V_{i} + 1_{j \neq i} \Sigma_{i})]: V_{i} \in \mathcal{T}_{\Sigma_{i}} \mathcal{P}(n_{i})\}.
\]
Let $\Sigma \in \mathcal{P}(\prod_{i = 1}^{d} n_{i})$, and $S_{1},S_{2} \in \mathcal{T}_{\Sigma}\mathcal{P}(\prod_{i = 1}^{d} n_{d})$. Define $\psi(\Sigma)$ to be the Boltzmann entropy of $\Sigma$
\begin{equation} \label{eq: Boltzmann entropy}
\psi(\Sigma) = \log(\vert \Sigma \vert) = tr(\log (\Sigma)).
\end{equation}
The Riemannian inner product induced by a normal likelihood was derived in \cite{moakher2011riemannian} as the Hessian of the Boltzmann Entropy
\begin{align} 
    g_{\Sigma}(S_{1}, S_{2}) &= - Hess \psi(\Sigma)(S_{1}, S_{2}) = -\frac{\partial^{2}}{\partial s \partial t} \psi(P + tS_{1} + sS_{2})\vert_{t = s = 0} \nonumber \quad t,s \in \mathbb{R}\\
    &= tr(\Sigma^{-1} S_{1} \Sigma^{-1} S_{2}) \label{eq: Riemannian metric full}.
\end{align}
By vectorizing the symmetric tangent vectors
\[
v(S) =  (S_{11}, S_{21}, \ldots, S_{d1}, S_{22}, \ldots, S_{d2}, \ldots, S_{dd}).
\]
The corresponding Riemannian gradient of a function defined on $\mathcal{P}(d)$, $f$, is given as
\begin{align} 
    vec(\nabla_{\Sigma}^{R} f) &= G^{-1}(\Sigma) v(\nabla_{\Sigma}^{E} f) \quad \text{for} \quad G^{-1}(\Sigma) = \Sigma \otimes \Sigma \label{eq: vectorized Riemannian gradient} \\
    \implies \nabla_{\Sigma}^{R} f &= \Sigma [\nabla_{\Sigma}^{E} f] \Sigma. \label{eq: Matrix Riemannian gradient}
\end{align}
Given a collection of Symmetric Positive Definite (SPD) matrices $\{\Sigma_{i}\}_{i=1}^{D}$, we could endow each of these independently with an affine-invariant metric to generate a structure known as a product manifold, which is essentially just the Cartesian product of the component manifolds. However, given $\Sigma = \otimes_{i = 1}^{D} \Sigma_{i}$ is itself an SPD matrix, one should ask whether we could naturally "induce" a geometry on each of the $\Sigma_{i}$ for $i \in \{1,\ldots,D\}$ by simply endowing $\otimes_{i = 1}^{D} \Sigma_{i}$ directly with an affine-invariant metric. Such a structure would be known as a pull-back metric.
\begin{definition}
    Let $\phi: \mathcal{M} \rightarrow \mathcal{N}$ be a smooth map between two manifolds $\mathcal{M}$ and $\mathcal{N}$. Suppose $\mathcal{N}$ has the additional structure of a metric $g$, and let $q \in \mathcal{M}$ with tangent vectors $v_{1}, v_{2} \in T_{q}M$. The corresponding pullback metric $\phi^{*} g$ is defined on $\mathcal{M}$ as $\phi^{*} g_{q}(v_{1},v_{2}) = g_{\phi(q)}(d\phi_{q} [v_{1}], d\phi_{q}[v_{2}])$.
\end{definition}
We will refer to the pullback metric in matrix form using corresponding upper case notation $\Phi^{*} G$. The pullback metric in full generality for the $D$-way case is described in the following result.
\begin{proposition} \label{prop: Non orthogonal AI metric}
    Let $\phi: \times_{i = 1}^{D} \mathcal{P}(d_{i}) \rightarrow \mathcal{P}^{\otimes}(\prod_{i = 1}^{D} d_{i})$. Endowing $\mathcal{P}^{\otimes}(\prod_{i = 1}^{D} d_{i})$ with the affine-invariant metric (\ref{eq: Affine Invariant metric}) yields the pullback metric $\Phi^{*} G$ on $\times_{i = 1}^{D} \mathcal{P}(d_{i})$ as the matrix:
\[
\Phi^{*} G = \left[
\begin{array}{cccc}
g_{11} & g_{12} & \cdots & g_{1n} \\
g_{21} & g_{22} & \cdots & g_{2n} \\
\vdots & \vdots & \ddots & \vdots \\
g_{n1} & g_{n2} & \cdots & g_{nn}
\end{array}
\right]
\]
where the entries are defined as
\[
 g_{ij} = \begin{cases}
[\prod_{k \not \in \{i,j\}} d_{k}] v(\Sigma_{i}^{-1}) v(\Sigma_{j}^{-1})^{T} & \text{if } i \neq j, \\
[\prod_{k \neq i} d_{k}] \Sigma_{i}^{-1} \otimes \Sigma_{i}^{-1} & \text{if } i = j.
\end{cases}
\]
\begin{proof}
    For the purpose of proof, we only pay attention to the case where $S_{1} = S_{2}$ to derive the norm of a tangent vector. In this case
    \begin{align}
    &\frac{\partial^{2}}{\partial t \partial s} log tr(\otimes_{i = 1}^{d} \Sigma_{i} + tS_{1} + s S_{2}) \nonumber \\
    &= tr\big[(\otimes_{j = 1}^{d} \Sigma_{j}^{-1})(\sum_{i = 1}^{d}[\otimes_{j = 1}^{d} (1_{j = i} V_{j} + 1_{j \neq i} \Sigma_{j})]) (\otimes_{j = 1}^{d} \Sigma_{j}^{-1}) (\sum_{i = 1}^{d}[\otimes_{j = 1}^{d} (1_{j = i} V_{j} + 1_{j \neq i} \Sigma_{j})])\big] \nonumber \\
    &= [\sum_{i = 1}^{d} tr(\Sigma_{i}^{-1} V_{i} \Sigma_{i}^{-1} V_{i}) \prod_{k/i} d_{k}] + 2*[\sum_{i = 1}^{d} \sum_{j \neq i} tr(\Sigma_{i}^{-1} V_{i}) tr(\Sigma_{j}^{-1} V_{j}) \prod_{k/\{i,j\}} d_{k}].
    \end{align}
\end{proof}
\end{proposition}
\begin{proposition}
    Let $\Phi^{*} G$ be the metric described in Proposition \ref{prop: Non orthogonal AI metric}, then $\Phi^{*} G$ is not positive definite.
    \begin{proof}
   For the matrix 
   \[
    \Phi^{*} G = \left[
    \begin{array}{cccc}
    g_{11} & g_{12} & \cdots & g_{1n} \\
    g_{21} & g_{22} & \cdots & g_{2n} \\
    \vdots & \vdots & \ddots & \vdots \\
    g_{n1} & g_{n2} & \cdots & g_{nn}
    \end{array}
    \right]
    \]
    where the entries are defined as
    \[
    g_{ij} = \begin{cases}
    [\prod_{k \not \in \{i,j\}} d_{k}] v(\Sigma_{i}^{-1}) v(\Sigma_{j}^{-1})^{T} & \text{if } i \neq j, \\
    [\prod_{k \neq i} d_{k}] \Sigma_{i}^{-1} \otimes \Sigma_{i}^{-1} & \text{if } i = j.
    \end{cases}
    \]
    Then partitioning $\Phi^{*} G$ as
    \[
    \Phi^{*}G = \left[
    \begin{array}{c|ccc}
    g_{11} & g_{12} & \cdots & g_{1n} \\
    \hline
    g_{21} & g_{22} & \cdots & g_{2n} \\
    \vdots & \vdots & \ddots & \vdots \\
    g_{n1} & g_{n2} & \cdots & g_{nn}
    \end{array}
    \right]
    \]
    then $\vert G \vert = \vert A \vert \vert R \vert$
    with 
    \begin{align} A &= g_{11} \\ 
    R &=  \left[ \begin{array}{ccc}
    g_{22} &  \cdots & g_{2n} \\
    \vdots & \ddots & \vdots \\
    g_{n2} & \cdots & g_{nn}
    \end{array} \right] - \left[ \begin{array}{c}
    g_{21} \\
    \vdots \\
    g_{n1}
    \end{array} \right] g_{11}^{-1} \left[ \begin{array}{ccc}
    g_{12} \cdots g_{1n}
    \end{array} \right].
    \end{align}
    Note that for any $i \neq j$,
    \begin{align}
        R_{i,j} &= g_{i,j} - g_{i,1} g_{1,1}^{-1} g_{1, j} \\
        &= [\prod_{k \not \in \{i,j\}} d_{k}] v(\Sigma_{i}^{-1}) v(\Sigma_{j}^{-1})^{T} - [\prod_{k \not \in \{i,1\}} d_{k}] v(\Sigma_{i}^{-1}) v(\Sigma_{1}^{-1})^{T} [\prod_{k \neq 1} \frac{1}{d_{k}}] \Sigma_{1} \otimes \Sigma_{1} [\prod_{k \not \in \{1,j\}} d_{k}] v(\Sigma_{1}^{-1}) v(\Sigma_{j}^{-1})^{T} \\
        &= [\prod_{k \not \in \{i,j\}} d_{k}] v(\Sigma_{i}^{-1}) v(\Sigma_{j}^{-1})^{T} - [\prod_{k \not \in \{1,i,j\}} d_{k}] v(\Sigma_{i}^{-1}) v(\Sigma_{1}^{-1})^{T} \Sigma_{1} \otimes \Sigma_{1} v(\Sigma_{1}^{-1}) v(\Sigma_{j}^{-1})^{T} = 0,
    \end{align}
    where the last line follows from
    \[
    [\prod_{k \not \in \{1,i,j\}} d_{k}] v(\Sigma_{i}^{-1}) v(\Sigma_{1}^{-1})^{T} \Sigma_{1} \otimes \Sigma_{1} v(\Sigma_{1}^{-1}) v(\Sigma_{j}^{-1})^{T} = [\prod_{k \not \in \{i,j\}} d_{k}] v(\Sigma_{i}^{-1}) v(\Sigma_{j}^{-1})^{T}.
    \]
    Hence, 
    \[
    \vert R \vert = \prod_{i = 2}^{n} \vert g_{ii} - g_{i1} g_{11}^{-1} g_{1i} \vert
    \]
    and observe $\vert g_{ii} - g_{i1} g_{11}^{-1} g_{1i} \vert = \vert g_{ii} \vert ( 1- 1) = 0$.
    \end{proof}
\end{proposition}

\begin{proposition} \label{prop: orthogonal AI metric}
    Under the orthogonalization condition $\vert \Sigma_{i} \vert = 1$ for all $i > 1$,
    \[
    \Phi^{*} G(\{ \Sigma_{i}\}_{i = 1}^{D}) = \oplus_{i = 1}^{D} G(\Sigma_{i}|\Sigma_{-1}) \quad \text{where} \quad G(\Sigma_{i}|\Sigma_{-1}) = \frac{1}{d_{-i}} \Sigma_{i}^{-1} \otimes \Sigma_{i}^{-1}
    \]
    and $\Phi^{*} G$ is positive definite.
    \begin{proof}
        First, note if $A$ and $B$ are positive definite matrices, the direct sum
        \[
        A \oplus B = \begin{pmatrix}
            A & 0 \\
            0 & B
        \end{pmatrix}
        \]
        is also positive definite. Recall from Proposition \ref{prop: Non orthogonal AI metric} the metric in algebraic form is:
        \begin{equation} \label{eq: metric tensor algebraic form}
        [\sum_{i = 1}^{d} tr(\Sigma_{i}^{-1} V_{i} \Sigma_{i}^{-1} V_{i}) \prod_{k/i} d_{k}] + 2*[\sum_{i = 1}^{d} \sum_{j \neq i} tr(\Sigma_{i}^{-1} V_{i}) tr(\Sigma_{j}^{-1} V_{j}) \prod_{k/\{i,j\}} d_{k}].
        \end{equation}
        Now note that the orthogonalization condition $\vert \Sigma_{i}\vert = 1$ is equivalent to the tangent space condition $tr([d\Sigma_{i}] \Sigma_{i}^{-1}) = 0$, which reduces (\ref{eq: metric tensor algebraic form}) to
        \[
        [\sum_{i = 1}^{d} tr(\Sigma_{i}^{-1} V_{i} \Sigma_{i}^{-1} V_{i}) \prod_{k/i} d_{k}]
        \]
        or in matrix form, $ \oplus_{i = 1}^{D} G(\Sigma_{i}).$
    \end{proof}
\end{proposition}

Let $V_{i} \in T_{\Sigma} P(d_{i})$, the tangent space constraint can be straightforwardly satisfied through the projection
\begin{equation} \label{eq: Projection operator}
P_{\Sigma_{i}}(V_{i}) = V_{i} - \frac{tr(V_{i} \Sigma^{-1})}{d_{i}} \Sigma_{i}.
\end{equation}
Letting $\vert \Sigma_{i} \vert = 1$ for $i > 1$, then the Riemannain gradient is given by
\begin{align*}
    \nabla_{\Sigma_{i}}^{R} &= (\prod_{j \neq 1} \frac{1}{d_{j}}) \Sigma_{1} \nabla_{\Sigma_{1}} \mathbb{E}_{q}[\log p(x,\theta) - \log q(\theta)] \Sigma_{1} \quad \text{ when } i = 1 \\
    &=  P_{\Sigma_{i}}([\prod_{j \neq i} \frac{1}{d_{j}}]\Sigma_{i} \nabla_{\Sigma_{i}} \mathbb{E}_{q}[\log p(x,\theta) - \log q(\theta)] \Sigma_{i} )\quad \text{otherwise}.
\end{align*}
Under this projection, the tangent space constraint will be naturally satisfied. Given a collection of Riemannian gradients on $\{\Sigma_{i}\}_{i = 1}^{D}$, the position updates are particularly easy to satisfy the following.
\begin{proposition}
If $\Sigma(t)$ is parameterized by a path in time as $\Sigma(t) = \otimes_{i = 1}^{D} \Sigma_{i}(t)$ for $t \in \mathbb{R}^{\geq 0}$, then under the orthogonalized metric of Proposition \ref{prop: orthogonal AI metric},
\begin{equation}
    \Sigma_{i}(t) = \exp_{\Sigma_{i}}tV_{i}(0) = \Sigma_{i}(0)^{\frac{1}{2}} \exp (t \Sigma_{i}(0)^{-\frac{1}{2}} V_{i}(0) \Sigma_{i}(0)^{-\frac{1}{2}}) \Sigma_{i}(0)^{\frac{1}{2}}.
\end{equation}
\begin{proof}
    First note if $\Sigma$ is parameterized by a path in time as $\Sigma(t) = \otimes_{i = 1}^{D} \Sigma_{i}(t)$ for $t \in \mathbb{R}^{+}$,
    \begin{align*}
V(0) = \frac{d}{dt} \Sigma(t) \vert_{t = 0} &= \sum_{i = 1}^{d}[\otimes_{j = 1}^{d}[1_{\{j = i\}}V_{j}(t) + 1_{\{j \neq i\}} \Sigma_{j}(t)] \vert_{t = 0} \\
&= \sum_{i = 1}^{d}[\otimes_{j = 1}^{d}[1_{\{j = i\}}V_{j}(0) + 1_{\{j \neq i\}} \Sigma_{j}(0)].
\end{align*}
    It then follows that
    \begin{align*}
        \Sigma(0)^{-\frac{1}{2}} V(0) \Sigma(0)^{-\frac{1}{2}} &= \sum_{i = 1}^{d} (\otimes_{i = 1}^{D} \Sigma_{i}(0)^{-\frac{1}{2}}) [\otimes_{j = 1}^{d}[1_{\{j = i\}}V_{j}(0) + 1_{\{j \neq i\}} \Sigma_{j}(t)] (\otimes_{i = 1}^{D} \Sigma_{i}(0)^{-\frac{1}{2}}) \\
        &= \sum_{i = 1}^{d}  \big[\otimes_{j = 1}^{D}\big( 1_{\{j = i\}} \Sigma_{j}(0)^{-\frac{1}{2}}V_{j}(0) \Sigma_{j}(0)^{-\frac{1}{2}} + 1_{j \neq i} I_{d_{j}}\big) \big].
    \end{align*}
    
     From  Section 2.3 of \cite{graham2018kronecker}, $\exp(A \otimes I) = \exp(A) \otimes I$, then it's straightforward to see that
    \begin{align*}
        \exp(t\Sigma(0)^{-\frac{1}{2}} V(0) \Sigma(0)^{-\frac{1}{2}}) &= \prod_{i = 1}^{D} \otimes_{j = 1}^{D} \exp(1_{\{j = i\}} \Sigma_{j}(0)^{-\frac{1}{2}}[t V_{j}(0)] \Sigma_{j}(0)^{-\frac{1}{2}} + 1_{j \neq i} I_{d_{j}}) \\
        &= \otimes_{i = 1}^{D} \exp(t \Sigma_{j}(0)^{-\frac{1}{2}}V_{j}(0) \Sigma_{j}(0)^{-\frac{1}{2}}).
    \end{align*}
    The result now follows immediately by observing $[(A \otimes B)(C \otimes D)]^{t} = A^{t}C^{t} \otimes B^{t}D^{t}$ for $t \in \mathbb{R}$.
\end{proof}

\end{proposition}

\section{Variational Bayes} \label{sec: Variational Bayes}
\subsection{Joint Kronecker Approximation} \label{sec: Structured Variational Approximation}
We will consider a variational approximation of the form $q(\Sigma \vert \nu, \{ A_{i} \}_{i = 1}^{D} ) \sim IW(\nu, \otimes_{i = 1}^{D} A_{i})$.  Letting $S = \sum_{i = 1}^{n} vec(\mathcal{X}_{i}) vec(\mathcal{X}_{i})^{T}$, the non-orthogonalized EBLO is derived in the Supplement as
\begin{align*}
 &LB(\nu_{v}, \{A_{i}\}_{i = 1}^{D}) = \mathbb{E}_{q}[\log P(X,\{\Sigma_{i}\}_{i = 1}^{D}) - \log q(\{\Sigma_{i}\}_{i = 1}^{D})] \\
 &= n\pi\prod_{i = 1}^{D} d_{i}  - \frac{\nu_{v}}{2} \big[tr([\otimes_{i = 1}^{D} A_{i}^{-1}](S + \Lambda)) - \prod_{i = 1}^{D} d_{i}] \\
 &+ (\nu_{v} - [n + \nu])\frac{\prod_{i = 1}^{D} d_{i}}{2} \log(2) + \log \Gamma_{\prod_{i = 1}^{D} d_{i}}(\frac{\nu_{v}}{2}) - \log \Gamma_{\prod_{i = 1}^{D} d_{i}}(\frac{\nu}{2}) + \frac{\nu}{2} \log \vert \Lambda \vert\\
 &+ \frac{\nu_{v} - [n + \nu]}{2}\big[\sum_{j = 1}^{D} d_{-j} log \vert A_{j} \vert - \prod_{i = 1}^{D} d_{i} \log(2)  - \sum_{i = 1}^{\prod_{i = 1}^{D} d_{i}} \psi(\frac{\nu_{v} - \prod_{i = 1}^{D} d_{i} + i}{2})\big].
\end{align*}
In the orthogonalized ELBO, we would instead have
\[
\sum_{j = 1}^{D} d_{-j} \log \vert A_{j} \vert = d_{-1} A_{1}
\]
which have corresponding gradients
\begin{align*}
    \nabla_{\nu_{v}}LB(\nu_{v}, \{A_{i}\}_{i = 1}^{D}) &= -\frac{1}{2} [tr(\otimes_{i = 1}^{D} A_{i}^{-1} [S + \Lambda]) - \prod_{i = 1}^{D} d_{i}] - \frac{n + \nu - \nu_{v}}{4} \sum_{i = 1}^{\prod_{i = 1}^{D} d_{i}} \psi(\frac{\nu_{v} - \prod_{i = 1}^{D} d_{i} + 1}{2}) \\
    \nabla_{A_{i}}LB(\nu_{v}, \{A_{i}\}_{i = 1}^{D}) &= \frac{\nu_{v}}{2} A_{i}^{-1} T^{i}(S + \Lambda, \Sigma_{-i}) A_{i}^{-1} - \frac{n + \nu}{2} d_{-1} A_{1}^{-1}
\end{align*}
where $T^{i}(S + \Lambda, \Sigma_{-i})$ was defined in Proposition \ref{prop: efficient trace}.

Note this then makes the Riemannian gradients of $\{A_{i}\}_{i = 1}^{D}$ particularly simple to compute as
\[
\nabla_{A_{p}}^{R} \mathbb{E}_{q}[p(\{\Sigma_{i}\}_{i = 1}^{d})] = \begin{cases}
     \prod_{j > 1}\frac{1}{d_{j}}[ T^{(1)}(S + \Lambda, \Sigma_{-1})  - \frac{ n + \nu}{2} d_{-1} A_{1} ] \quad \text{ if } p = 1 \\
     [\prod_{j \neq p} \frac{1}{d_{j}}] P_{A_{p}}(T^{(p)}(S + \Lambda, \Sigma_{-1})) \quad   \text{otherwise},
\end{cases}
\]
where $P_{A_{p}}(\cdot)$ was the projection operator defined in (\ref{eq: Projection operator}).

To deal with the optimization of the degrees of freedom, note that the gradient of $\nu_{V}$ is bottlenecked by the trace term. A priori computing the alternative representation
\begin{align*}
    - \frac{1}{2} tr(\otimes_{i = 1}^{D} A_{i} [S + \Lambda]) = -\frac{1}{2} tr(A_{i}^{-1} T^{(1)}(S + \Lambda, \Sigma_{-1}))
\end{align*}
and passing $ T^{(1)}(S + \Lambda, \Sigma_{-1})$ to the gradient of $A_{1}$ then allows $\nu_{v}$ to be updated with minimal computational overhead. To maintain the constraint $\eta_{v} > \prod_{i = 1}^{D} d_{i} + 1$, we substitute $\eta_{v} = \exp(z) + \prod_{i = 1}^{D} d_{i} + 1$ for $z \in \mathbb{R}$.
Gradients with respect to $z$ are then
\[
\nabla_{z} ELBO = [\frac{\prod_{i =1}^{D} d_{i}}{2}  + (n + \nu - \nu_{v}) \sum_{i = 1}^{\prod_{i = 1}^{D} d_{i}} \Psi'(\frac{\nu_{v} - i + 1}{2}) - \frac{1}{2} tr( A_{1}^{-1} T^{(1)}(S + \Lambda, \Sigma_{-1})) ]\exp(z) \label{eq: z gradient},
\]
wherein updates to $\eta_{v}$ are constructed as follows
\begin{itemize}
    \item $z = \log(\eta_{v} - \prod_{i = 1}^{D} d_{i} - 1)$
    \item Update $z^{*} \leftarrow z + \nabla_{z} ELBO$
    \item $\eta_{v}^{*} \leftarrow \eta_{v} + \exp(z^{*})$.
\end{itemize}

The algorithm for any variational algorithm discussed is described in Algorithm 2. %\ref{alg: SVB}#%  
A subtle note of this algorithm is that step \ref{step:normalization} is not entirely necessary, as in theory the geodesic flow should retain the determinant equal to one constraint for any given time step. However, we found it useful to ensure that no computational round-off will accumulate error in this regard.

\begin{algorithm} \label{alg: SVB}
\caption{Joint Multiway Variational Bayes}
\begin{algorithmic}[1]
\STATE \textbf{Precomputation:}
\STATE \textbf{Input:} $S = \sum_{i = 1}^{N} y_{i} y_{i}^{T}$, mode dimensions $\mathcal{T} = \{d_{1}, \ldots, d_{D}\}$.
\FOR{each \( m \) in \( \{1,\ldots, D\} \)}
    \STATE Compute \( S_{\Gamma[m]} \) iteratively according to proposition \ref{prop: efficient trace}.
    \ENDFOR
\STATE \textbf{Output: } $S_{\Gamma[m]}$ for $m \in \{1,\ldots, D\}$.
\STATE \textbf{Main Algorithm:}
\STATE \textbf{Input: } $\{ S_{\Gamma}^{(m)}, A_{i} \}_{m = 1}^{D}$, $\nu_{v}$, $t \in \mathbb{R}$, number of iterations, $n_{t}$
\FOR{$i \in 1,\ldots, n_{t}$}
    \STATE $\theta = \{ \nu_{v}, \{A_{j}\}_{j = 1}^{D}\}$
    \STATE Compute $T^{(1)}(S, \Sigma_{-1})$ according to proposition \ref{prop: efficient trace}, let $tr(\otimes_{m = 1}^{D} \Sigma_{m}^{-1} S) = tr(\Sigma_{1}^{-1} T^{(1)})$.
    \STATE With $z = \log(\nu_{V} - \prod_{m = 1}^{D} d_{m} - 1)$, compute $g_{z}^{E} = \nabla_{z} \mathbb{E}_{q}[p(\theta, X \vert T^{(1)}(S,\Sigma_{-1})) - q(\theta)]$.
    \STATE Set $\nu_{V} \leftarrow \nu_{v} + \exp(z + g_{z}^{E})$.
    \FOR{each \( m \) in \( \{1,\ldots, D\} \)}
        \STATE Compute $g_{m} = \nabla_{A_{m}} \mathbb{E}_{q}[p(\theta, X) - q(\theta)]$
        \IF{$m \neq 1$}
        \STATE Set $g_{m}^{R} \leftarrow P_{A_{m}}([\prod_{j = 1, j \neq m}^{D} \frac{1}{d_{j}}]A_{m} g_{m} A_{m})$
        \ELSE
        \STATE $g_{1}^{R} = [\prod_{j = 1, j \neq 1}^{D} \frac{1}{d_{j}}]A_{1} g_{1} A_{1} $
        \ENDIF
    \ENDFOR
    \FOR{each \( m \) in \( \{1,\ldots, D\} \)}
        \STATE Compute $A_{m}(t) = \exp_{A_{m}}(t g_{m}) = A_{m}^{\frac{1}{2}} \exp(t A_{m}^{-\frac{1}{2}} g_{m}^{R} A_{m}^{-\frac{1}{2}}) A_{m}^{\frac{1}{2}}$
        \STATE Set $A_{m} \leftarrow A_{m}(t)$
        \IF{$m > 1$}
        \STATE Set $A_{m} \leftarrow \frac{A_{m}}{\vert A_{m} \vert^{1/d_{m}}}$ \label{step:normalization}
        \ENDIF
    \ENDFOR
\ENDFOR
\STATE \FOR{\( i \in 1:n_{it}\)} 
\STATE \textbf{Return} \( \{ A_{j}\}_{j = 1}^{D} \).
\ENDFOR
\end{algorithmic}
\end{algorithm}

\subsection{Mean Field Approximation} \label{sec: Mean Field Approximation}
The mean field approximation is one which admits a factored density as
\[
q(\{\Sigma_{i}\}_{i = 1}^{D} \vert \{\nu_{i}, A_{i}\}_{i = 1}^{D}) = \prod_{i = 1}^{D} q(\Sigma_{i} \vert \nu_{v_{i}}, A_{i}).
\]
According to the model of \cite{hoff2011separable}, letting $Y_{i} = vec(\mathcal{Y}_{i})$, the Bayesian model is specified as
\begin{align*}
    y_{i} &\sim \mathcal{N}(0, \otimes_{i = D}^{1} \Sigma_{i})\\
    \Sigma_{i} &\sim IW(d_{i} + 2, \frac{\gamma^{1/D}}{d_{i}} I_{d_{i}}) \quad \text{for} \quad \gamma = tr(\hat{S}).
\end{align*}
The mean field approximation to the model would then be identified by specifying independent distributions on $\{ \Sigma_{i}\}_{i = 1}^{D}$ according to $q(\Sigma_{i}) \sim IW(\nu_{v_{i}}, A_{i})$ and as shown in the Supplement, yields the lower bound for the mean field approximation
\begin{align*}
 &LB_{MF}(\nu_{v}, \{A_{i}\}_{i = 1}^{D}) = \mathbb{E}_{q_{MF}}[\log P(X,\{\Sigma_{i}\}_{i = 1}^{D}) - \log q(\{\Sigma_{i}\}_{i = 1}^{D})] \\
 &= \frac{n}{2}\pi\prod_{i = 1}^{D} d_{i}  -  \big[tr([\otimes_{i = 1}^{D} A_{i}^{-1}][\prod_{j = 1}^{D} \frac{\nu_{v_{j}}}{2} S]) + \sum_{j = 1}^{D} \nu_{v_{j}} \frac{\gamma^{\frac{1}{D}}}{d_{i}}tr(A_{i}^{-1}) - \sum_{i = 1}^{D} \frac{\nu_{v_{i}}d_{i}}{2}] \\
 &+ \sum_{i = 1}^{D} \big[(\nu_{v_{i}} - [nd_{-i} + \nu_{i}])\frac{d_{i}}{2} \log(2) + \log \Gamma_{d_{i}}(\frac{\nu_{v_{i}}}{2}) - \log \Gamma_{d_{i}}(\frac{\nu_{i}}{2}) + \frac{1}{2} [\nu_{i}d_{i} \log (\frac{\gamma^{(1/D)}}{d_{i}}) - \nu_{v_{i}}\log \vert A_{i} \vert] \\
 &+ \frac{\nu_{v_{i}} - [nd_{-i} + \nu_{i}]}{2}\big[ log \vert A_{i} \vert -  d_{i} \log(2)  - \sum_{i = 1}^{ d_{i}} \psi(\frac{\nu_{v_{i}} -  d_{i} + 1}{2})\big] \big].
\end{align*}
Which has corresponding Euclidean gradients given by
\begin{align*}
\nabla_{A_{p}} LB(\{ \nu_{v_{j}}\}_{j = 1}^{D}, \{\Sigma_{i}\}_{i = 1}^{D}) &= A_{p}^{-1} T^{(p)}(\prod_{j = 1}^{D} \frac{\nu_{v_{j}}}{2}S + \sum_{j = 1}^{D} \nu_{v_{j}} \frac{\gamma^{\frac{1}{D}}{d_{i}}}{2} I_{d_{i}}, \Sigma_{-i}) A_{p}^{-1} - \frac{nd_{-i} + \nu_{i}}{2} A_{i}^{-1} \\
\nabla_{\nu_{v_{i}}} LB(\{ \nu_{v_{j}}\}_{j = 1}^{D}, \{\Sigma_{i}\}_{i = 1}^{D}) &= [\frac{( d_{i} + 1) - d_{-i}(n + \nu + \prod_{j = 1}^{D} d_{j} + 1)}{2}] A_{p} + \frac{\prod_{j = 1}^{D} \nu_{j} }{2} C^{p}(S + \Lambda).
\end{align*}
We note the only modifications that need to be made to Algorithm 2 %\ref{alg: SVB}%
to employ the mean field approximation are the change in ELBO, step \ref{step:normalization} is entirely omitted, and steps 16-20 may instead be combined into a single step as $g_{i}^{R} \leftarrow A_{i} g_{i} A_{i}$. This change in ELBO obviously requires optimizing several degrees of freedom $\{\nu_{v_{i}}\}_{i = 1}^{D}$ rather than a single variational degree of freedom. One could optimize each of these successively conditional on the previous; however, we found quicker convergence by optimizing them simultaneously.

\section{Empirical Comparisons} \label{sec: Empirical Comparisons}
\subsection{Simulated data examples}

In this section, we illustrate the performance of the joint Kronecker structured variational approximation of the unstructured IW posterior versus the mean field approximation of the Bayesian model with independent priors.

\begin{figure}[h!]  
    \centering
    % Row 1
    \begin{minipage}{0.45\textwidth}
        \centering
        \includegraphics[width=\textwidth]{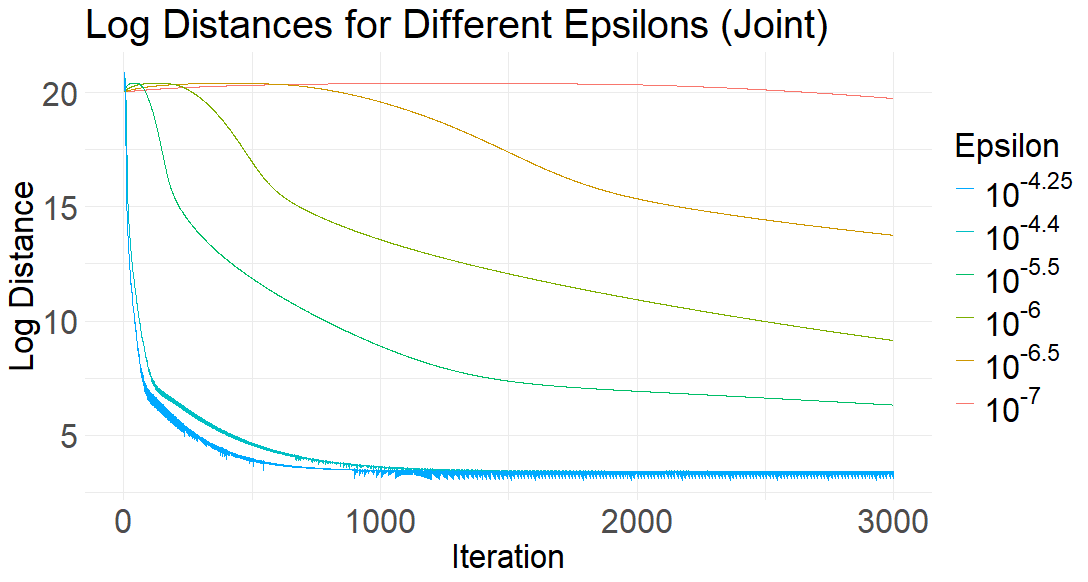}
    \end{minipage}\hfill
    \begin{minipage}{0.45\textwidth}
        \centering
        \includegraphics[width=\textwidth]{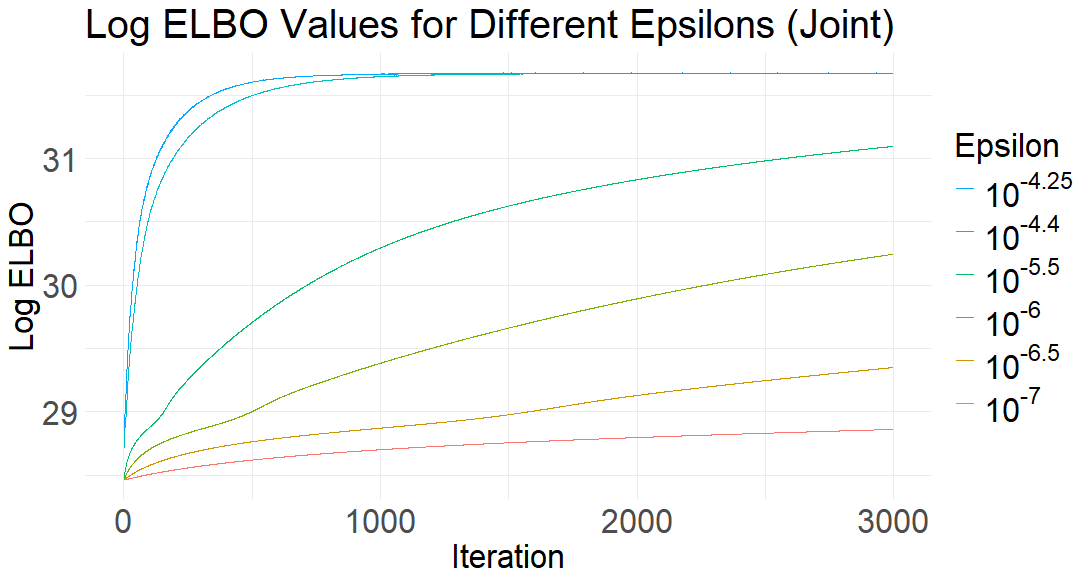}
    \end{minipage}
    
    \vspace{0.5cm} % Space between rows

    % Row 2
    \begin{minipage}{0.45\textwidth}
        \centering
        \includegraphics[width=\textwidth]{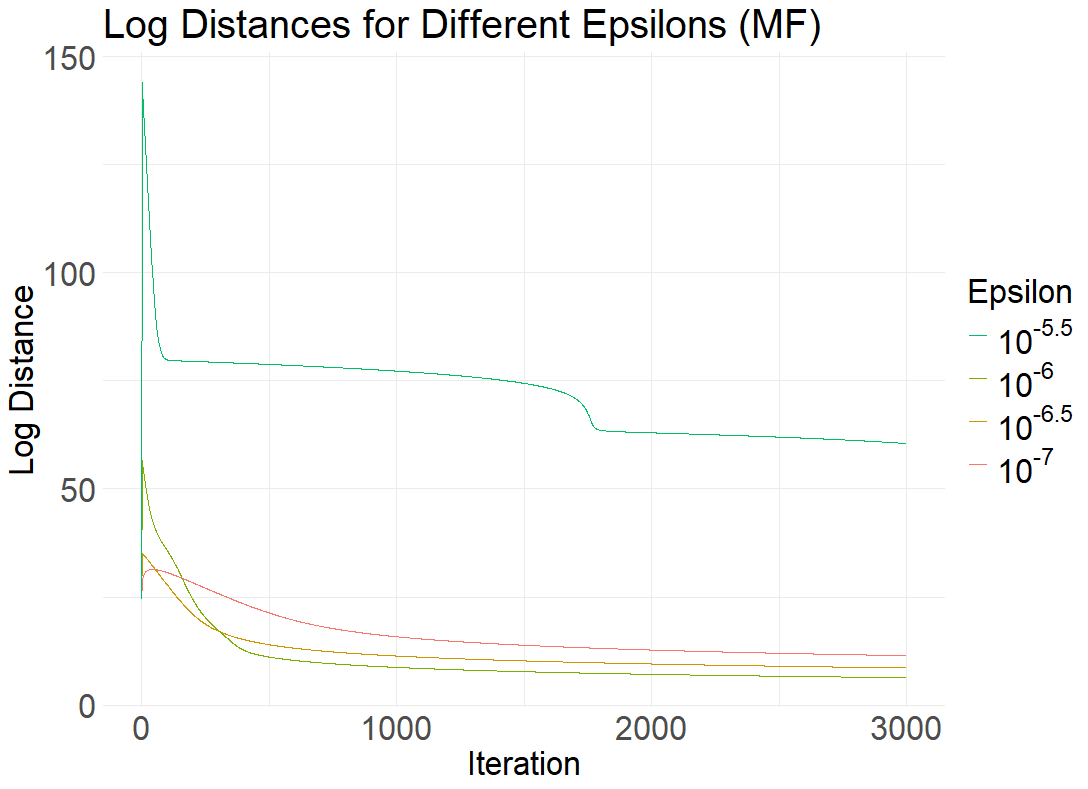}
    \end{minipage}\hfill
    \begin{minipage}{0.45\textwidth}
        \centering
        \includegraphics[width=\textwidth]{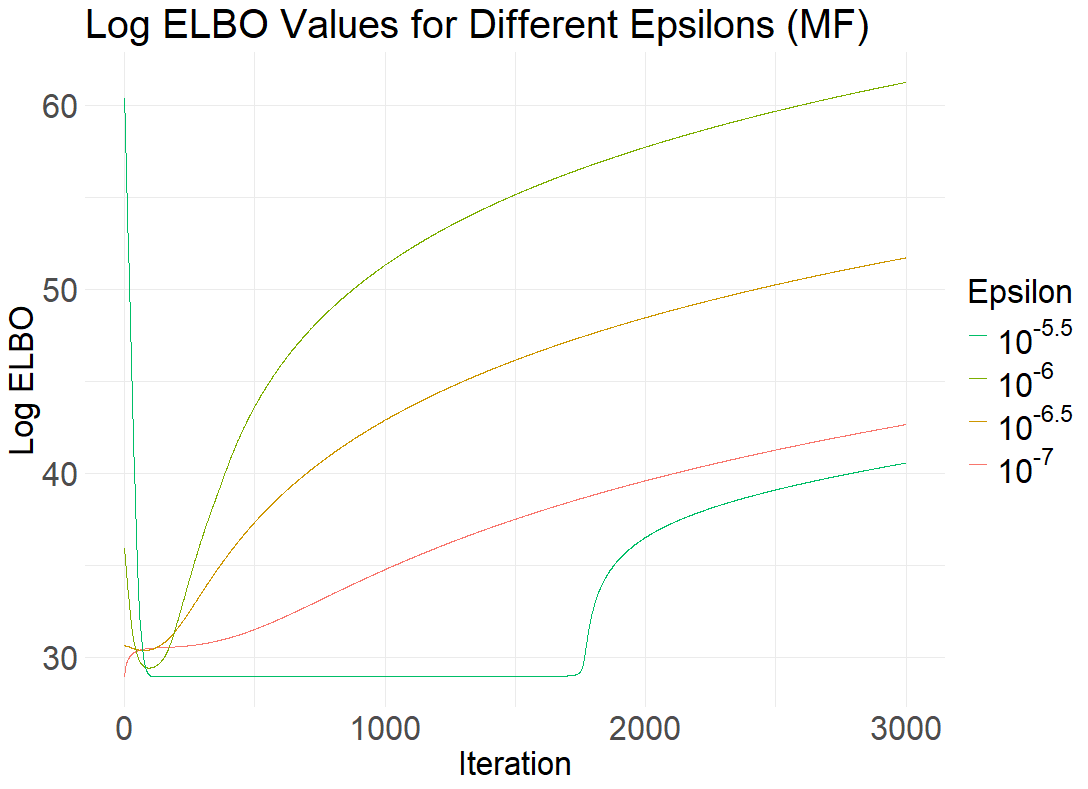}
    \end{minipage}

    \caption{Log distances (distance here is defined as the sum of squares error $\|\mathbb{E}_{Q}[\Sigma \vert Y] - \Sigma \|_{F}^{2}$) and log ELBO for the joint approximation using the pullback metric (first row), and the mean field approximation using the product manifold metric (second row). Observations are colored according to the joint stepsize used for updates to both $\{A_{i}\}_{i = 1}^{D}$ and the degrees of freedom updates ($\nu_{v}$ for the joint approximation and $\nu_{v_{i}}$ for the mean field approximation).}
    \label{fig:log distances and elbos mf vs joint}
\end{figure}

We will make comparisons of the geometric and model choices in terms of iterations to convergence. In our first generated data example, we generated $\Sigma_{i} \in \mathcal{P}(d_{i})$ with $d = (5,6,4,3)$, with $y_{1},\ldots, y_{n} \sim N(0, \otimes_{i = 1}^{4} \Sigma_{i})$. In Figure \ref{fig:log distances and elbos mf vs joint}, we demonstrate the time to convergence for the joint variational approximation of Section \ref{sec: Structured Variational Approximation} for the unstructured Inverse Wishart and the mean field approximation of  the Bayesian model in \cite{hoff2011separable}. We compare the time to convergence using a pullback metric for the joint approximation and a product manifold metric for the mean field approximation, using a global step-size shared between all scale matrices $\{A_{i}\}_{i = 1}^{D}$ and degrees of freedom ($\nu_{v}$ for the joint approximation and $\{\nu_{v_{i}}\})_{i = 1}^{D}$ for the mean field approximation). We emphasize here that this isn't meant to be a comparison of which geometric choice is superior from an optimization perspective, but instead to emphasize that the mean field approximation introduces several nuisance parameters which substantially limit optimization speed in comparison to the joint approximation. 

Taking the most successful of these global $\epsilon$ choices ($\epsilon_{joint} = 10^{-4.4}$ and $\epsilon_{MF} = 10^{-6}$, respectively), we iterated VB, using log distance of the joint and mean field approximation to ground truth to assess convergence, we then illustrate their variational predictive performance through average Mahalanobis distance. That is, let $\{\hat{A}_{i}\}_{i = 1}^{D}, \hat{\nu}_{v}$ be the parameters at convergence for the joint model, or $\{\hat{A}_{i}\}_{i = 1}^{D}, \{\hat{\nu}_{i}\}_{i = 1}^{D}$ for the Mean field model . In the joint model, we generate
\begin{align*}
    \Sigma^{(1)}, \ldots, \Sigma^{(K)} &\sim IW(\hat{\nu}_{v}, \otimes_{i = 1}^{D} \hat{A}_{i})\\
    y_{1}^{(t)}, \ldots, y_{m}^{(t)} &\sim N(0, \Sigma^{(t)}) \\
    M^{(t)} &= \frac{1}{m} \sum_{i = 1}^{m} y_{i}^{(t)} \Sigma^{-1} y_{i}^{(t)}.
\end{align*}
Or in the mean field model as
\begin{align*}
    \Sigma_{i}^{(t)} &\sim IW(\hat{\nu}_{i}, \hat{A}_{i}), \quad i \in \{1,\ldots, D\}, \quad t \in \{1,\ldots, K\} \\
    y_{1}^{(t)}, \ldots, y_{m}^{(t)} &\sim N(0, \otimes_{i = 1}^{D} \Sigma_{i}^{(t)}) \\
    M^{(t)} &= \frac{1}{m} \sum_{i = 1}^{m} y_{i}^{(t)} \Sigma^{-1} y_{i}^{(t)}.
\end{align*}
In each simulated data example, $m = 100$, and $K = 200$
In each figure, we let separable VB run for 3000 iterations for the joint variational approximation, and 10000 for the mean field approximation. We note that we do not necessarily run the mean field approximation until convergence, as these figures are meant to demonstrate the computational efficiency of the joint model over the mean field model. 

In Figure \ref{fig:log distances and elbos mf vs joint}  for large epsilon, the pullback metric for the joint approximation remains stable and converges to the optimum in less than 1000 iterations.  Compared to others, the mean field approximation never converged for any epsilon choice in 3000 iterations and was unstable for large epsilon (it could not evaluate a single iteration beyond the choice of $\epsilon_{MF} = 10^{-5.5}$ whereas $\epsilon_{Joint}$ was successfully able to evaluate for all choices up to $\epsilon_{Joint} = 10^{-4.25}$. We note that while the joint approximation looks unstable for the largest choice of $\epsilon_{Joint}$, this is due to $\epsilon_{Joint}$ also controlling the step size for the degrees of freedom updates, which causes oscillatory behavior).

\begin{figure}[ht]
    \centering
    \includegraphics[width=0.48\textwidth, height=0.2\textheight]{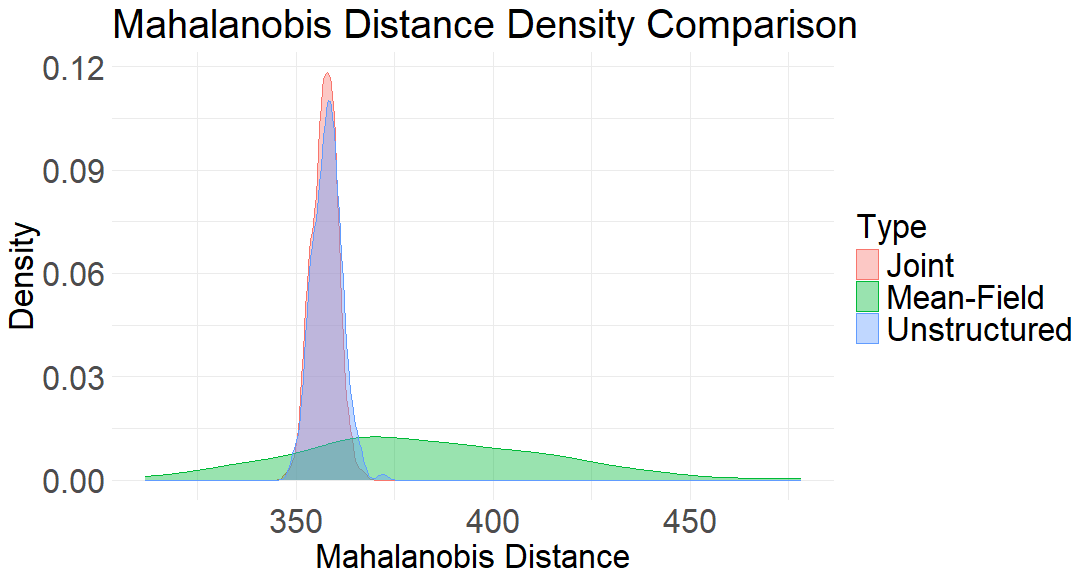}
    \hfill
    \includegraphics[width=0.48\textwidth, height=0.2\textheight]{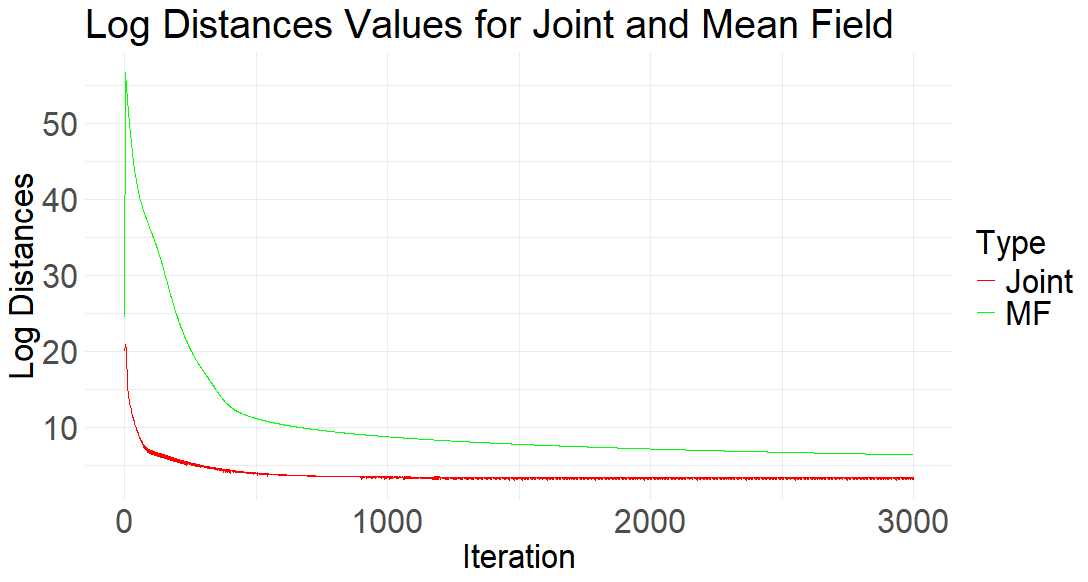}
    
    % Global caption for the entire figure layout
    \label{fig:Mahalanobis comparisons}
    \caption{On the left: Posterior predictive Mahalanobis distance for unstructured IW posterior (blue), the Joint Kronecker Variational Approximation (red), and the Mean Field Approximation to the Bayesian Model with independent priors (green). On the right: Sum of squares distances from the ground truth to the mean of $\Sigma$ under the joint variational distribution (red line) and mean field approximation of the independent prior model (green line).
    The joint approximation clearly provides a better approximation of the unstructured posterior than the mean-field approximation of the larger Bayesian model. This should however be unsurprising, as a mean field approximation of a Bayesian model with independent priors should naturally lead to higher variance samples. More interestingly, we see the speed of convergence for the joint approximation is substantially quicker for the joint approximation than for the mean field approximation. We note that this isn't necessarily a result of the pullback metric used during optimization, as even the product manifold based optimization was efficient in comparison to optimization of the mean field approximation.}
    \label{fig: posterior predictive Mahalanobis distance ground truth separable}
\end{figure}

Figure \ref{fig: posterior predictive Mahalanobis distance ground truth separable} compares the average Mahalanobis distances when $\Sigma$ is truly separable as $\Sigma = \otimes_{i = 1}^{D} \Sigma_{i}$.  It is evident that while all 3 center in a similar region, the mean field approximation of the independent prior model has variances that act multiplicatively, indicating a lack of fit to the true model. On the right side of Figure \ref{fig: posterior predictive Mahalanobis distance ground truth separable} clearly while both models converge to approximately the sum of squares distance from the ground truth, the joint model is able to converge in a fraction of the time to the mean field approximation.

\begin{figure}[h!]
    \centering
    % Row 1
    \begin{minipage}{0.45\textwidth}
        \centering
        \includegraphics[width=\textwidth, height = .2\textheight]{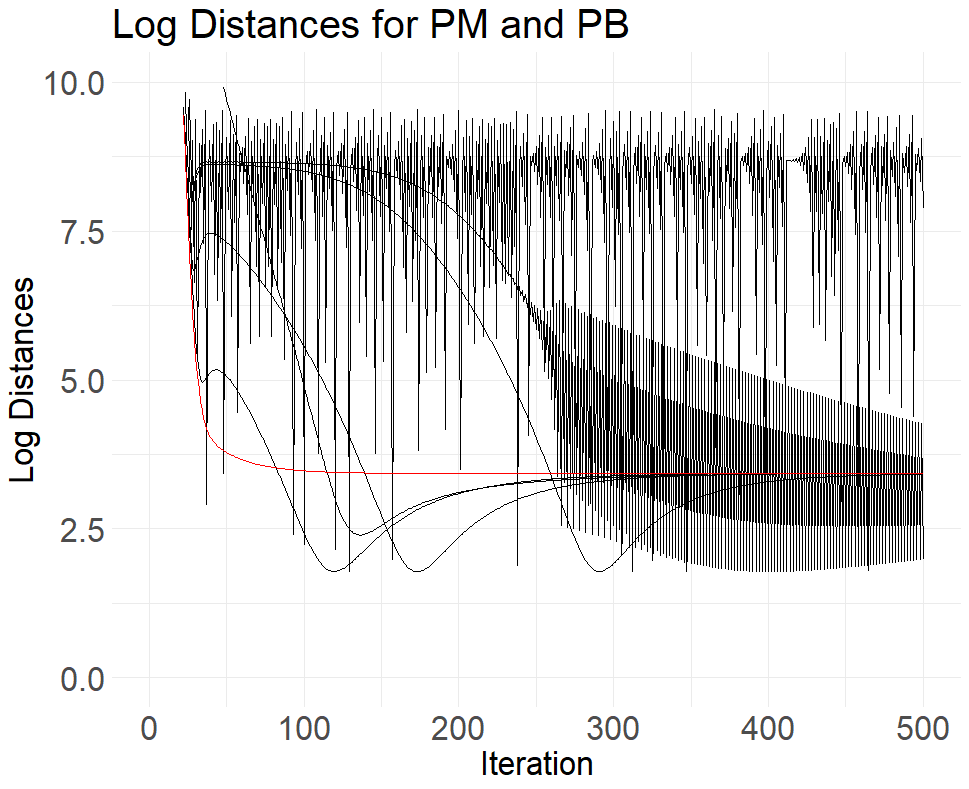}
    \end{minipage}
    \hfill
    \begin{minipage}{0.45\textwidth}
        \centering
        \includegraphics[width=\textwidth, height = .2\textheight]{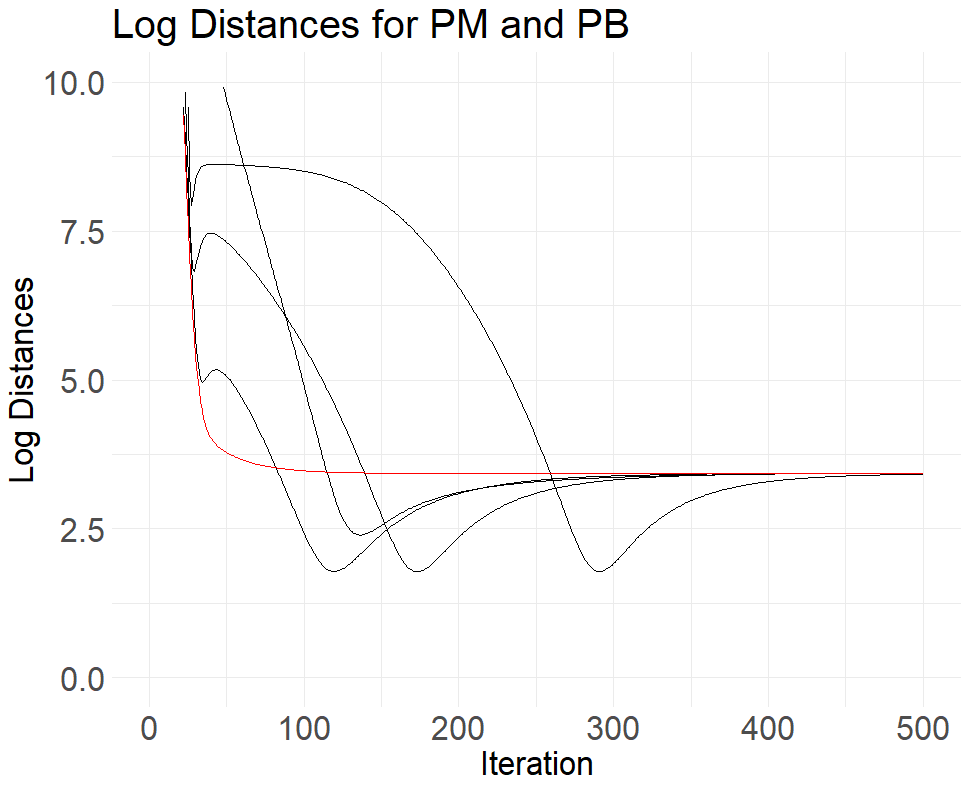}
    \end{minipage}
    \caption{Log distances of mean to ground truth $(\| \mathbb{E}_{q}[\Sigma] - \otimes_{i = 1}^{D} \Sigma_{i} \|_{F}^{2})$ for pullback metric (red line) and product metric (black lines) under the largest stable (choices) for stepsizes for scale matrices. Step size for the degrees of freedom was fixed at $\epsilon_{\nu_{v}} = 10^{-5}$ for comparibility. The figure on the left includes all stepsizes of $\epsilon_{\{A_{i}\}_{i =1}^{D}}^{(PM)} \in \{10^{-4.9}, 10^{-5}, 10^{-5.1}, 10^{-5.35}, 10^{-5.5}, 10^{-6} \}$ and $\epsilon_{\{A_{i}\}_{i =1}^{D}}^{(PB)} = 10^{-3.5}$. This is mainly to emphasize that for larger choices of $\epsilon^{(PM)}$, the optimization destabilizes, yielding the oscillatory patterns without improvement in convergence. Removing the first two values of $\epsilon_{\{A_{i}\}_{i =1}^{D}}^{(PM)} \in \{10^{-4.9}, 10^{-5}\}$ to denoise the left figure yields the figure on the right. Note that the pullback metric reaches the point of convergence substantially quicker than any of the step size choice under the product manifold.}
    \label{fig:log distances pullback vs product}
\end{figure}

In Figure \ref{fig:log distances pullback vs product}, we compare the iterations with convergence for the joint approximation under the product manifold geometry and the pullback geometry. It is of note that it is not optimal to compare these metrics using a global stepsize shared between $\nu_{v}$ and $\{A_{i}\}_{i = 1}^{D}$. Instead in this setting, we used separate step sizes for the updates to $\nu_{v}$ and $\{A_{i}\}_{i = 1}^{D}$, with the former using $\epsilon_{\nu_{v}} = 10^{-5}$ and the latter using the largest stable step-size for the respective metric, which was $\epsilon_{\{A_{i}\}_{i =1}^{D}}^{(PB)} = 10^{-3.5}$ for the pullback metric, and $\epsilon_{\{A_{i}\}_{i =1}^{D}}^{(PM)} \in \{10^{-4.9}, 10^{-5}, 10^{-5.1}, 10^{-5.35}, 10^{-5.5}, 10^{-6} \}$ for the product manifold.

Lastly, in Table \ref{table: time to convergence}, we compare the number of iterations to convergence in terms of sum of squares error as $\arg \min_{i} \| \hat{\Sigma}_{i} - \hat{\Sigma}_{N}\|_{F} < \beta$ when the likelihood model is misspecified and instead we have a true covariance which is a mutliway way covariance perturbed by low rank noise, $\Sigma = \otimes_{i = 1}^{D} \Sigma_{i} + \sum_{i = 1}^{r} x_{i} x_{i}^{T}$. These numbers in Table \ref{table: time to convergence} illustrate that the joint model is robust in terms of time to convergence of the model under likelihood misspecification, whereas the mean field model is not.

\clearpage

\begin{table}[h!]
\centering
\begin{tabular}{|c|c|c|c|c|}
\hline
Setting & r = 1 & r = 3 & r = 5 & r = 10 \\ \hline
Joint      & 303       &  730      &  933       & 1220       \\ \hline
Mean Field      & 5946     & 14513       & 29972       & $>$30000       \\ \hline
\end{tabular}
\caption{Iterations to convergence in terms of sum of squared error to maximal value, $\arg \min_{i} \| \hat{\Sigma}_{i} - \hat{\Sigma}_{N} \|_{F} < \beta$ for misspecified models where the true covariance was specified as $\Sigma = \otimes_{i = 1}^{D} \Sigma_{i} + \sum_{i = 1}^{r} x_{i}x_{i}^{T}$ with $x_{i} \sim N(0, \xi I_{d})$. Here $\beta = .005$, $\xi = .2$,  $N = 3000$ for the joint approximation and $N = 30000$ for the mean field approximation with $\epsilon_{joint} = 10^{-5}$ and $\epsilon_{MF} = 10^{-7}$ being the largest epsilon choices stable across all choices of $r$.}
\label{table: time to convergence}
\end{table}

\subsection{Global Trade Data Example}
In this section we analyze the subset of UN trade commodity data considered in \cite{hoff2011separable}. Our data is $\mathcal{A} \in \mathbb{R}^{30 \times 30 \times 6 \times 10}$ consisting of 10 temporal slices composed of {\em Exporting nation, importing nation, commodity type} in $\mathbb{R}^{30}, \mathbb{R}^{30}, \mathbb{R}^{6}$, respectively. We treat the model as vectorized temporal observations $y_{1}, \ldots, y_{10} \in \mathbb{R}^{6 \times 30 \times 30}$. Given the short term nature of the time series, we assume there will be minimal temporal dependence in the joint posterior of $(\mu, \Sigma)$, and treat these as a time independent normal Inverse-Wishart (IW) prior:
\begin{align*}
    y_{1}, \ldots, y_{10} &\sim N_{5400}(\mu, \Sigma)\\
    \mu \vert \mu_{0},  \Sigma &\sim N(\mu \vert \mu_{0}, \Sigma) \\
    \Sigma \vert \Psi, \nu &\sim IW(\Sigma \vert \Psi, \nu).
\end{align*}
Such that $\mu_{0} = \bar{y}$, $\Psi = C$, and $\nu = \prod_{i = 1}^{D} d_{i} + 2$, where $C$ is the sample covariance matrix. For direct analysis of $\Sigma$, we marginalize out $\mu$ to yield the corresponding marginal posterior on $\Sigma$ (see \cite{degroot2005optimal} for further details):
\begin{align*}
\Sigma &\vert \Psi, \nu, Y \sim IW(\Sigma, \Psi + S) \\
S &= \sum_{i = 1}^{10} (y_{i} - \bar{y}) (y_{i} - \bar{y})^{T}.
\end{align*}
Now proceed by fitting the joint Kronecker variational Bayes model as $\Sigma \vert Y \approx Q_{\nu_{v}, \{A_{i}\}_{i = 1}^{D}} (\Sigma) \sim IW(\nu_{v}, \otimes_{i = 3}^{1} A_{i}).$  Each $A_{i}$ was initialized randomly with $A_{i} \sim IW(d_{i} + 2, \frac{\gamma}{d_{i}}I_{d_{i}})$ with $\gamma = 5$, and $\nu_{v} = \prod_{i = 1}^{3} d_{i} + 2$ was our initial value for $\nu_{V}$.

\begin{figure}
    \centering % Centers the figure

      % First image
      \includegraphics[width=0.95\textwidth]{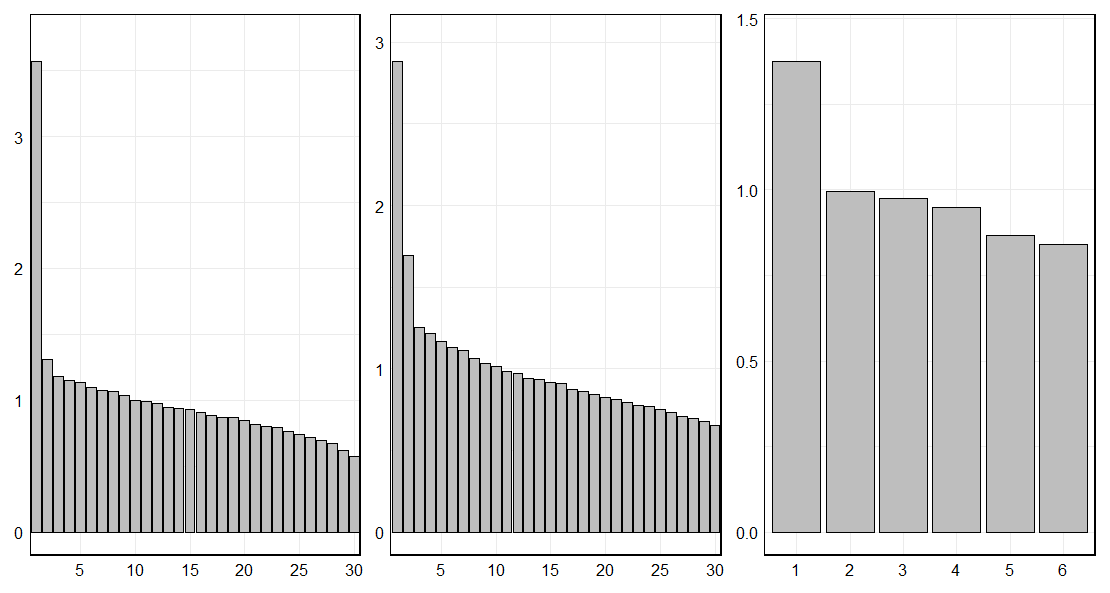} % Adjust the path and image file
      \vspace{1cm} % Adds vertical space between the images
    
      % Second image
      \includegraphics[width=0.95\textwidth]{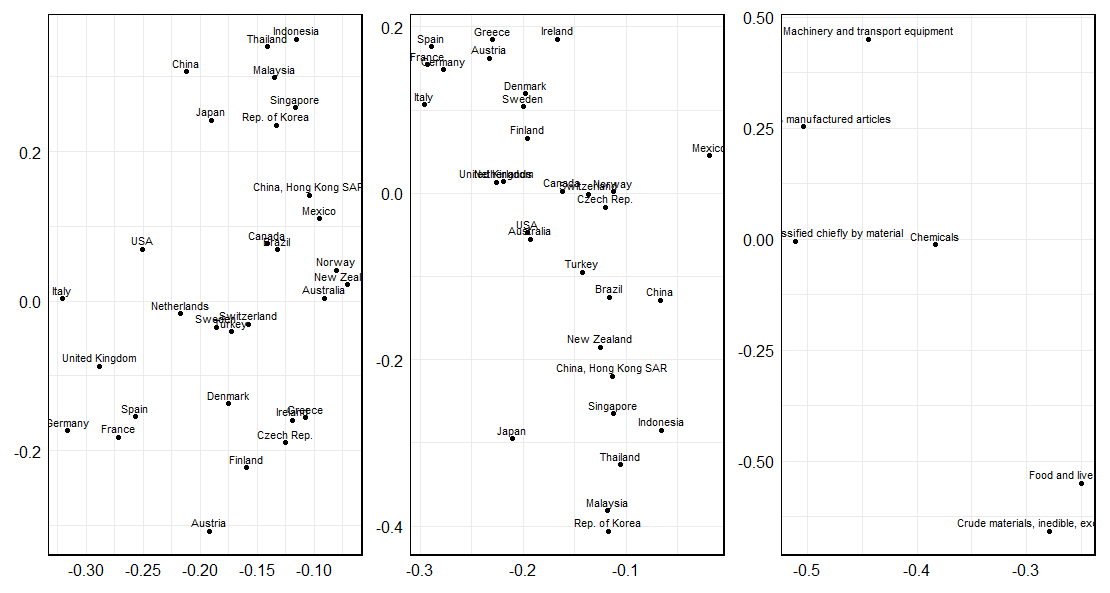} % Adjust the path and image file
    \caption{Mean correlation matrices corresponding to $\Sigma_{1}$, $\Sigma_{2}$ and $\Sigma_{3}$ at convergence. The top row are the eigenvalues of the corresponding correlation matrices, the bottom plot is of the first two components of the eigenvectors (the first component is on the x-axis, and the second component on the y-axis). }
    \label{fig:eigenvector eigenvalue plots}
\end{figure}

At convergence of Algorithm 2, %\ref{alg: SVB}%
we compute the eigendecomposition of the mean correlation matrices corresponding to $\Sigma_{1}$, $\Sigma_{2}$ and $\Sigma_{3}$ corresponding to exporting nations, importing nations, and commodity type, respectively. These results are found in Figure \ref{fig:eigenvector eigenvalue plots}. Compared to \cite{hoff2011separable}, we found an identical eigenvalue structure. Regarding the eigenvectors of the correlation matrices, we observed a similar clustering pattern of the first two components where geographic location was the dominating factor for clustering in the first 2 correlation matrices and the extent to which a commodity type is a finished good was the primary contribution for clustering in the last correlation matrix. In particular for $\Sigma_{1}$, moving from the top right of the plot to the bottom left, the trend generally goes from east Asian to western European countries, with intermediate countries (USA, Canada, etc.) in the middle. Note that the clustering appears more pronounced for the Asian countries than European countries. For $\Sigma_{2}$, the same trend can be observed when going from the bottom right to the top left, although the clustering effect appears to be more pronounced for the European countries than for the Asian countries. For the Eigenvalue plot of $\Sigma_{3}$, the clustering of the eigenvectors for commodities seems determined by the quality of finish for the product. For example: machinery, manufactured, chemicals, etc. in the top left as a "finished" good, and in the bottom right crude materials, food (potentially raw)/live animals clustered together. This shows the same potential for modeling the mode-wise covariances with a factor analytic structure as discussed in \cite{hoff2011separable}.

\section{Conclusions}
In this work we presented a technique for geometrically informed optimization of variational Bayesian approximations for posteriors of mutliway covariances. The mean field approximation, while typically employed for simplicity, demonstrates obvious difficulties for convergence in comparison to using a joint approximation of the multiway scale matrix. Regarding differential geometric optimization, we demonstrated the utility of the pullback metric in endowing the space with a product manifold geometry as a robust technique for efficient optimization on the space of Kronecker-structured SPD matrices.

For future work, \cite{lin2019riemannian} give a metric on Cholesky factors of SPD matrices, and analyze its properties through its pushforward to SPD space. Letting $\mathcal{C}(d)$ denote the space of Cholesky factors of $d\times d$ SPD matrices, this metric was developed through the observation that any $L \in \mathcal{C}(d)$ can be described through
\[
\mathcal{L}(\Sigma) = \mathbb{D}(L) + \lfloor L \rfloor
\]
where $L = \mathcal{L}(\Sigma)$ is the Cholesky factor of $\Sigma$ , $\mathbb{D}(\cdot)$ is the diagonal component of a matrix, and $\lfloor \cdot \rfloor$ is the strictly lower triangular component. They observe the topological manifold structure of this space and construct a metric on this space as:
\[
g_{L}(V,W) = \langle \lfloor V \rfloor, \lfloor W \rfloor \rangle_{F} + <\mathbb{D}(L)^{-1} \mathbb{D}(V), \mathbb{D}(L)^{-1} \mathbb{D}(W) \rangle_{F},\quad V,W \in T_{L}
\]
and derive the corresponding geodesic connecting $L_{0} = \mathcal{L}(P_{0})$ to $L_{1} = \mathcal{L}(P_{1})$ as: 
\begin{align*}
    &\lfloor L_{0} \rfloor + t(\lfloor L_{1} \rfloor - \lfloor L_{0} \rfloor) + \mathbb{D}(L_{0}) \exp (t \mathbb{D}(L_{0}) \log (\mathbb{D}(L_{0})^{-1} \mathbb{D}(L_{1})) \mathbb{D}(L_{0})^{-1}) \\
    &= \lfloor L_{0} \rfloor + t(\lfloor L_{1} \rfloor - \lfloor L_{0} \rfloor) + \mathbb{D}(L_{1})^{t} \mathbb{D}(L_{0})^{1-t}
\end{align*}
We note however these geodesics do not yield a geodesically convex optimization problem for the Bartlett decomposition of an Inverse-Wishart, although is conditionally geodesically convex. While yielding a potential loss of wall clock efficiency, we may still be interested in considering such a metric in the case where we have array data where any individual mode may be exceptionally large such as in \cite{dorta2018training}, where a Vecchia-like approximation on the scale matrix of a Bartlett distribution may provide substantially better computational scalability over the use of the Kronecker SPD manifold endowed with the pullback of the affine-invariant metric. The key observation to see this is to note that the log Cholesky metric can preserve structured strictly lower triangular sparsity, and noting that the affine-invariant metric cannot make use of structured sparsity.

Outside of methodological improvements, it should be of note that the mean field approximation for this problem is known to be limited according to to the work of \cite{pitsanis1997kronecker}, who showed the best Kronecker product approximation of a general dense matrix $S$ under the Frobenius norm is limited
\[
\min_{A,B} \|S - A \otimes B\|_{F} \geq \sigma_{2} 
\]
where $\sigma_{2}$ is the second singular value of a reshaped version of $S$. We note from this that samples under the mean field approximation will always maintain this bound, as mean field samples will always be perfectly separable. However, while the mean of the joint approximation will maintain this bound, general samples will not, as even with a separable scale matrix, separable SPD matrices still form a measure 0 set on the support of an Inverse-Wishart distribution.






\section{Supplementary Material}
\section*{Proof of the ELBO}
\subsection{Joint Approximation}
With  a variational approximation of the form
\[
q(\Sigma \vert \nu, \{ A_{i} \}_{i = 1}^{D} ) \sim IW(\nu, \otimes_{i = 1}^{D} A_{i})
\]
Letting $S = \sum_{i = 1}^{n} vec(\mathcal{X}_{i}) vec(\mathcal{X}_{i})^{T}$, observing $\Sigma^{-1} \sim W(\nu, \otimes_{i = 1}^{D} A_{i}^{-1})$, wherein $\mathbb{E}[\otimes_{i = 1}^{D} \Sigma_{i}^{-1}] = \nu \otimes_{i = 1}^{D} A_{i}^{-1}$, then per equation 34 of \cite{braun2010variational}:
\begin{equation} \label{eq: log determinant expectation}
    \mathbb{E}_{q}[- log \vert \otimes_{i = 1}^{D} \Sigma_{i} \vert] = -\sum_{i = 1}^{D} d_{-i} log \vert A_{i} \vert - \prod_{i = 1}^{D} d_{i} \log(2)  - \sum_{i = 1}^{\prod_{i = 1}^{D} d_{i}} \psi(\frac{\nu - \prod_{i = 1}^{D} d_{i} + i}{2}) .
\end{equation}
First observe by linearity of the trace
\begin{equation} \label{eq: trace expectation}
    \mathbb{E}_{q}[tr(-\frac{1}{2} \otimes_{i = 1}^{D} \Sigma_{i}^{-1} S)] = tr(-\frac{\nu}{2} \otimes_{i = 1}^{D} A_{i}^{-1} S)
\end{equation}
then the expecation of the log likelihood with respect to Q is given by
\begin{equation} \label{eq: Likelihood expectation}
    \mathbb{E}_{q}[\mathcal{L}(y;\{\Sigma_{i}\}_{i = 1}^{D})] =  n(\prod_{i = 1}^{D} d_{i}\pi - \frac{1}{2} \big[d_{-1}log \vert A_{1} \vert - \prod_{i = 1}^{D} d_{i} \log(2)  - \sum_{i = 1}^{\prod_{i = 1}^{D} d_{i}} \psi(\frac{\nu - \prod_{i = 1}^{D} d_{i} + i}{2}) \big]) -\frac{\nu_{v}}{2} tr(\otimes_{i = 1}^{D} A_{i}^{-1} S) 
\end{equation}
Likewise, the expectation with respect to the log prior is given by
\begin{align*}
    \mathbb{E}_{q}[p(\{\Sigma_{i}\}_{i = 1}^{d})] &= \frac{\nu}{2} \log \vert \Lambda \vert - \frac{\nu \prod_{i = 1}^{D} d_{i}}{2} \log (2) - \log \Gamma_{\prod_{i = 1}^{D} d_{i}}(\frac{\nu}{2}) \\
    &- \frac{\nu + \prod_{i = 1}^{d} d_{i} + 1}{2} \big[d_{-1} log \vert A_{1} \vert - \prod_{i = 1}^{D} d_{i} \log(2)  - \sum_{i = 1}^{\prod_{i = 1}^{D} d_{i}} \psi(\frac{\nu_{v} - \prod_{i = 1}^{D} d_{i} + i}{2})\big] \\
    &- \frac{\nu_{v}}{2} tr(\Lambda \otimes_{i = 1}^{D} A_{i}^{-1}).
\end{align*}

The negative expectation under the variational distribution is:
\begin{align*}
\mathbb{E}_{q}[-\log q(\{\Sigma_{i}\}_{i = 1}^{D})] &= -\frac{\nu_{v} d_{-1}}{2} \log \vert A_{1} \vert + \frac{\nu_{v} \prod_{i = 1}^{D} d_{i}}{2} \log (2) + \log \Gamma_{\prod_{i = 1}^{D} d_{i}}(\frac{\nu_{v}}{2}) \\
    &+ \frac{\nu_{v} + \prod_{i = 1}^{d} d_{i} + 1}{2} \big[d_{-1} log \vert A_{1} \vert + \prod_{i = 1}^{D} d_{i} \log(2)  + \sum_{i = 1}^{\prod_{i = 1}^{D} d_{i}} \psi(\frac{\nu_{v} - \prod_{i = 1}^{D} d_{i} + i}{2})\big] \\
    &+ \frac{\nu_{v} \prod_{i = 1}^{D} d_{i}}{2}.
\end{align*}
Note here we used the fact
\[
\mathbb{E}_{q}[-\frac{1}{2} tr([\otimes_{i = 1}^{D} \Sigma_{i}^{-1}] [\otimes_{i = 1}^{D} A_{i}])] = -\frac{1}{2} tr(\mathbb{E}_{q}[\otimes_{i = 1}^{D} \Sigma_{i}^{-1}][\otimes_{i = 1}^{D} A_{i}]) = -\frac{\nu \prod_{i = 1}^{D} d_{i}}{2}
\]

Combining these terms yields the non-orthogonalized EBLO
\begin{align*}
 &LB(\nu_{v}, \{A_{i}\}_{i = 1}^{D}) = \mathbb{E}_{q}[\log P(X,\{\Sigma_{i}\}_{i = 1}^{D}) - \log q(\{\Sigma_{i}\}_{i = 1}^{D})] \\
 &= n\pi\prod_{i = 1}^{D} d_{i}  - \frac{\nu_{v}}{2} \big[tr([\otimes_{i = 1}^{D} A_{i}^{-1}](S + \Lambda)) - \prod_{i = 1}^{D} d_{i}] \\
 &+ (\nu_{v} - [n + \nu])\frac{\prod_{i = 1}^{D} d_{i}}{2} \log(2) + \log \Gamma_{\prod_{i = 1}^{D} d_{i}}(\frac{\nu_{v}}{2}) - \log \Gamma_{\prod_{i = 1}^{D} d_{i}}(\frac{\nu}{2}) + \frac{\nu}{2} \log \vert \Lambda \vert\\
 &+ \frac{\nu_{v} - [n + \nu]}{2}\big[\sum_{j = 1}^{D} d_{-j} log \vert A_{j} \vert - \prod_{i = 1}^{D} d_{i} \log(2)  - \sum_{i = 1}^{\prod_{i = 1}^{D} d_{i}} \psi(\frac{\nu_{v} - \prod_{i = 1}^{D} d_{i} + i}{2})\big].
\end{align*}
Where in the orthogonalized ELBO, we would instead have
\[
\sum_{j = 1}^{D} d_{-j} \log \vert A_{j} \vert = d_{-1} A_{1}.
\]

\subsection{Mean Field Approximation}

With a factored variational distribution
\[
q(\{\Sigma_{i}\}_{i = 1}^{D} \vert \{\nu_{i}, A_{i}\}_{i = 1}^{D}) = \prod_{i = 1}^{D} q(\Sigma_{i} \vert \nu_{v_{i}}, A_{i})
\]
And the Bayes model of \cite{hoff2011separable}, letting $Y_{i} = vec(\mathcal{Y}_{i})$:
\begin{align*}
    y_{i} &\sim \mathcal{N}(0, \otimes_{i = D}^{1} \Sigma_{i})\\
    \Sigma_{i} &\sim IW(d_{i} + 2, \frac{\gamma^{1/D}}{d_{i}} I_{d_{i}})
\end{align*}
Where $\gamma = tr(\hat{S})$.

Then using equation \ref{eq: log determinant expectation},
\[
\mathbb{E}_{q}[\log \vert \otimes_{i = 1}^{D} \Sigma_{i} \vert] = \sum_{i = 1}^{D} d_{-i} \log \mathbb{E}_{q}[\vert \Sigma_{i} \vert] = \sum_{i = 1}^{D} d_{-i} [\log \vert A_{i} \vert - d_{i} \log (2) - \sum_{i = 1}^{d_{i}} \psi(\frac{\nu_{v_{i}} - d_{i} + i}{2})]
\]

The corresponding expectations are
\begin{align*}
    \mathbb{E}_{q}[\log L(Y;\{\Sigma_{i}\}_{i = 1}^{D})] = \frac{n}{2}[ \pi \prod_{i = 1}^{D} d_{i} - \sum_{i = 1}^{D} d_{-i} [\log \vert A_{i} \vert - d_{i} \log(2) - \sum_{i = 1}^{d_{i}}\psi(\frac{\nu_{v_{i}} - d_{i} + i}{2})]] - \frac{\prod_{j = 1}^{D} \nu_{v_{j}}}{2} tr(\otimes_{i = 1}^{D} A_{i}^{-1} S)
\end{align*}

\begin{align*}
    \mathbb{E}_{q}[p(\{\Sigma_{i}\}_{i = 1}^{d})] &= \sum_{i = 1}^{D}\big[ \frac{\nu_{i}}{2} d_{i} \log (\frac{\gamma^{(1/D)}}{d_{i}}) - \frac{\nu_{i} d_{i}}{2} \log (2) - \log \Gamma_{ d_{i}}(\frac{\nu_{i}}{2}) \\
    &- \frac{\nu_{i} +  d_{i} + 1}{2} \big[log \vert A_{1} \vert -  d_{i} \log(2)  - \sum_{i = 1}^{ d_{i}} \psi(\frac{\nu_{v_{i}} -  d_{i} + i}{2})\big] \\
    &- \frac{\nu_{v_{i}} \frac{\gamma^{1/D}}{d_{i}}}{2} tr( A_{i}^{-1}) \big]
\end{align*}

\begin{align*}
\mathbb{E}_{q}[-\log q(\{\Sigma_{i}\}_{i = 1}^{D})] &= \sum_{i = 1}^{D} \big[ -\frac{\nu_{v_{i}}}{2} \log \vert A_{i} \vert + \frac{\nu_{v_{i}} d_{i}}{2} \log (2) + \log \Gamma_{ d_{i}}(\frac{\nu_{v_{i}}}{2}) \\
    &+ \frac{\nu_{v_{i}} + d_{i} + 1}{2} \big[ log \vert A_{i} \vert +  d_{i} \log(2)  + \sum_{i = 1}^{ d_{i}} \psi(\frac{\nu_{v_{i}} -  d_{i} + i}{2})\big] \\
    &+ \frac{\nu_{v_{i}}  d_{i}}{2} \big]
\end{align*}
which yield the non-orthogonalized EBLO
\begin{align*}
 &LB(\nu_{v}, \{A_{i}\}_{i = 1}^{D}) = \mathbb{E}_{q}[\log P(X,\{\Sigma_{i}\}_{i = 1}^{D}) - \log q(\{\Sigma_{i}\}_{i = 1}^{D})] \\
 &= n\pi\prod_{i = 1}^{D} d_{i}  - \frac{\nu_{v}}{2} \big[tr([\otimes_{i = 1}^{D} A_{i}^{-1}](S + \Lambda)) - \prod_{i = 1}^{D} d_{i}] \\
 &+ (\nu_{v} - [n + \nu])\frac{\prod_{i = 1}^{D} d_{i}}{2} \log(2) + \log \Gamma_{\prod_{i = 1}^{D} d_{i}}(\frac{\nu_{v}}{2}) - \log \Gamma_{\prod_{i = 1}^{D} d_{i}}(\frac{\nu}{2}) + \frac{\nu}{2} \log \vert \Lambda \vert\\
 &+ \frac{\nu_{v} - [n + \nu]}{2}\big[\sum_{j = 1}^{D} d_{-j} log \vert A_{j} \vert - \prod_{i = 1}^{D} d_{i} \log(2)  - \sum_{i = 1}^{\prod_{i = 1}^{D} d_{i}} \psi(\frac{\nu_{v} - \prod_{i = 1}^{D} d_{i} + i}{2})\big]
\end{align*}



%%%%%%%%%%%%%%%%%%%%%%%%%%%%%%%%%%%%%%%%%%%%%%
%% Funding information, if any,             %%
%% should be provided in the                %%
%% funding section.                         %%
%%%%%%%%%%%%%%%%%%%%%%%%%%%%%%%%%%%%%%%%%%%%%%
\noindent
\textbf{Funding:} MTW acknowledges partial support from NIH awards R01GM135926 and 1T32HD113301-01.

%The second author was supported in part by NIH Grant ???????????.

\bibliographystyle{plain}
\bibliography{sample-base}

\end{document}